\documentclass[useAMS,usenatbib]{mn2e}
\usepackage{amssymb,amsmath,overpic,dcolumn,apjfonts}
\usepackage{color,transparent}
\usepackage[usenames,dvipsnames,svgnames,table]{xcolor}

\allowdisplaybreaks[1]
\newcolumntype{.}{D{.}{.}{-1}}

\renewcommand{\vec}[1]{\mathbf{#1}}
\newcommand{\svec}[1]{\boldsymbol{#1}}
\newcommand{\mr}[1]{\mathrm{#1}}

\newcommand{\propsiminn}[2]{\mathrel{\vcenter{
  \offinterlineskip\halign{\hfil$##$\cr
    #1\propto\cr\noalign{\kern2pt}#1\sim\cr\noalign{\kern-2pt}}}}}

\newcommand{\propsim}{\mathpalette\propsiminn\relax}

\renewcommand{\epsilon}{\varepsilon}
\newcommand{\grad}{\svec{\nabla}}
\newcommand{\vecxi}{\svec{\xi}}

\newcommand{\eq}[1]{equation~\eqref{#1}}
\newcommand{\eqp}[1]{equation~\ref{#1}}
\newcommand{\Eq}[1]{Equation~\eqref{#1}}
\newcommand{\ta}[1]{Table~\ref{#1}}
\newcommand{\fig}[1]{Figure~\ref{#1}}
\newcommand{\se}[1]{\S~\ref{#1}}
\newcommand{\app}[1]{Appendix~\ref{#1}}

\newcommand{\detun}{\delta\omega}

\newcommand{\Omegsp}{\Omega_\mr{spin}}
\newcommand{\Omegspd}{\dot{\Omega}_\mr{spin}}

\newcommand{\omegdyn}{\omega_*}
\newcommand{\Omegdiff}{\Omega-\Omegsp}
\newcommand{\dOmeg}{\delta\Omega}
\newcommand{\Omegd}{\dot{\Omega}}

\newcommand{\tdyn}{t_*}
\newcommand{\ttherm}{t_\mr{th}}
\newcommand{\tsync}{t_\mr{sync}}
\newcommand{\tgw}{t_\mr{gw}}
\newcommand{\tinsp}{t_\mr{merge}}
\newcommand{\tgroup}{t_\mr{group}}
\newcommand{\tcool}{t_\mr{cool}}
\newcommand{\teddy}{t_\mr{eddy}}

\newcommand{\Eorb}{E_\mr{orb}}
\newcommand{\Eth}{E_\mr{th}}

\newcommand{\rl}{\mr{rl}}
\newcommand{\trl}{\mr{trl}}

\newcommand{\Porb}{P_\mr{orb}}
\newcommand{\Prl}{P_\rl}
\newcommand{\Ptrl}{P_\trl}

\newcommand{\Is}{I_*}

\newcommand{\Q}{\mathcal{Q}}
\newcommand{\Qt}{\Q_\mr{t}}
\newcommand{\betas}{\beta_*}

\newcommand\delad{\nabla_\mr{ad}}

\newcommand{\omegt}{\sigma}

\newcommand{\torq}{\tau}
\newcommand{\torqtrav}{\torq_\mr{trav}}
\newcommand{\torqstand}{\torq_\mr{stand}}
\newcommand{\torqtrans}{\torq}

\newcommand{\Edtide}{\dot{E}_\mr{tide}}
\newcommand{\Edheat}{\dot{E}_\mr{heat}}
\newcommand{\Etide}{E_\mr{tide}}
\newcommand{\Es}{E_*}

\newcommand{\Msun}{M_\sun}
\newcommand{\Rsun}{R_\sun}

\newcommand{\krxirabs}{|k_r \xir |}
\newcommand{\krxirm}{\krxirabs_\mr{max}}

\newcommand{\Teff}{T_\mr{eff}}

\newcommand{\Gamcrys}{\Gamma_\mr{crys}}

\newcommand{\tA}{t_\mr{A}}
\newcommand{\vA}{v_\mr{A}}

\newcommand{\brunt}{Brunt-V\"{a}is\"{a}l\"{a}}

\newcommand{\xir}{\xi_r}
\newcommand{\h}{\mr{h}}
\newcommand{\xih}{\xi_\h}

\newcommand{\coid}[1]{\texttt{CO#1}}
\newcommand{\heid}[1]{\texttt{He#1}}

\newcommand{\co}{carbon/oxygen}

\newcommand{\wdw}{WD}
\newcommand{\wdws}{WDs}
\newcommand{\Wdw}{WD}

\newcommand{\cowd}{\co{} \wdw}
\newcommand{\hewd}{helium \wdw}

\renewcommand{\sun}{\odot}

\newcommand\altaffilmark[1]{$^{#1}$}
\newcommand\altaffiltext[1]{$^{#1}$}
\newcommand{\acknowledgments}{\begin{small}
    \section*{Acknowledgments}\end{small}}

\title{Tidal resonance locks in inspiraling white dwarf binaries}

\author[Burkart et al.]{
\parbox[t]{\textwidth}{
Joshua Burkart,\altaffilmark{1}
Eliot Quataert,\altaffilmark{1,2}
Phil Arras,\altaffilmark{3} and
Nevin N. Weinberg\altaffilmark{4}}
\vspace*{6pt} \\
\altaffiltext{1}{Department of Physics, 366 LeConte Hall, University of California,
Berkeley, CA\ \ 94720, USA}\\
\altaffiltext{2}{Department of Astronomy \& Theoretical Astrophysics
  Center, 601 Campbell Hall, University of California Berkeley, CA\ \ 94720, USA}\\
\altaffiltext{3}{Department of Astronomy, University of Virginia,
P.O. Box 400325, Charlottesville, VA\ \ 22904-4325, USA}\\
\altaffiltext{4}{Department of Physics and MIT Kavli Institute, MIT, 77 Massachusetts Avenue,
Cambridge, MA~~02139, USA}
}

\date{Accepted to MNRAS}

\begin{document}
\maketitle
\label{firstpage}

\begin{abstract}
We calculate the tidal response of helium and carbon/oxygen (C/O) white dwarf (WD) binaries inspiraling due to gravitational wave emission. We show that resonance locks, previously considered in binaries with an early-type star, occur universally in WD binaries. In a resonance lock, the orbital and spin frequencies evolve in lockstep, so that the tidal forcing frequency is approximately constant and a particular normal mode remains resonant, producing efficient tidal dissipation and nearly synchronous rotation. We show that analogous locks between the spin and orbital frequencies can occur not only with global standing modes, but even when damping is so efficient that the resonant tidal response becomes a traveling wave. We derive simple analytic formulas for the tidal quality factor $\Qt$ and tidal heating rate during a g-mode resonance lock, and verify our results numerically. We find that $\Qt\sim10^7$ for orbital periods $\lesssim1$ -- 2~hr in C/O WDs, and $\Qt\sim10^9$ for $\Porb\lesssim3$ -- 10~hr in helium WDs. Typically tidal heating occurs sufficiently close to the surface that the energy should be observable as surface emission. Moreover, near an orbital period of $\sim10$~min, the tidal heating rate reaches $\sim 10^{-2} L_\sun$, rivaling the luminosities of our fiducial WD models. Recent observations of the 13-minute double-WD binary J0651 are roughly consistent with our theoretical predictions.  Tides naturally tend to generate differential rotation; however, we show that the fossil magnetic field strength of a typical WD can maintain solid-body rotation down to at least $\Porb\sim10$~min even in the presence of a tidal torque concentrated near the WD surface.
\end{abstract}

\begin{keywords}
binaries, white dwarfs, waves, instabilities
\end{keywords}

\section{Introduction}
In this work, we consider the effect of tides in detached white dwarf (\wdw{}) binaries  inspiraling due to energy and angular momentum loss by gravitational waves. Our analysis is motivated by several important questions. For example, to what degree should short-period \wdw{} binaries exhibit synchronized rotational and orbital motion? Should \wdws{} in close binaries be systematically hotter than their isolated counterparts, as a result of tidal dissipation? What is the thermal state of \wdws{} prior to the onset of mass transfer?

Several past studies have applied linear perturbation theory to the  problem considered in this work. \citet{campbell84} and \citet{iben98} applied the theory of the equilibrium tide to \wdw{} binaries, using parameterized viscosities to estimate the tidal torque. \citet{willems10} considered turbulent convective damping acting on the equilibrium tide, as originally considered by \citet{zahn77} for late-type stars, and showed that this effect is not able to synchronize a \wdw{} binary within its gravitational wave inspiral time.

\citet{rathore05} and \citet{fuller11} moved beyond the large-scale, nonresonant equilibrium tide, and considered the tidal excitation of standing g-modes during  inspiral, analyzing the behavior of wave amplitudes as a system sweeps through resonances. However, neither study allowed the \wdw{}'s spin rate to evolve, an assumption that eliminated the physics highlighted in this work.

In this paper we also focus on tidally excited g-modes in \wdw{} binaries; one of our goals is to assess whether the resonantly excited  ``dynamical tide'' represents a traveling or standing wave. This amounts to whether a tidally generated wave is able to reflect at its inner and outer radial turning points. If reflection cannot occur, then a damping time of order the group travel time across the mode propagation cavity results; if reflection does occur, then the wave amplitude can build up significantly during close resonances between g-mode frequencies and the tidal forcing frequency.

This question has been addressed before in the context of main-sequence stars. \citet{zahn75} employed a traveling wave description of the dynamical tide in the context of early-type stars with radiative envelopes, assuming waves would be absorbed near the surface by rapid radiative diffusion. \citet{goldreich89} enhanced this argument, showing that dynamical tides first cause tidal synchronization in such a star's outer regions, leading to the development of critical layers and even stronger radiative damping. However, they did not assess whether angular momentum redistribution could enforce solid-body rotation and thereby eliminate critical layers; we address this important point in \se{s:reslockval}. In the absence of critical layers, \citet{witte99} introduced the phenomenon of resonance locks (\se{s:reslock}), which rely on the large wave amplitudes produced during eigenmode resonances. We will show that similar resonance locks occur ubiquitously in close \wdw{} binaries.

\citet{goodman98} considered the case of late-type stars with convective envelopes, and showed that tidally generated waves excited at the edge of the convection zone steepen and break near the cores of such stars. \citet{fuller12}, in their study of the tidal evolution of \wdw{} binaries, found that the dynamical tide in a \cowd{} instead breaks near the outer turning point. As such, they invoked an outgoing-wave boundary condition  in their analysis. We find that their assumption may not be generally applicable due to an overestimate of the degree of wave breaking; see \se{s:wbreak}. As a result the dynamical tide may represent a standing wave for a substantial portion of a \wdw{} binary's inspiral epoch.

This paper is organized as follows. In \se{s:prelim} we provide pertinent background information on \wdws{} and tidal effects that our subsequent results rely on. In \se{s:dynregimes} we give a broad overview of the results we derive in more detail in \S\S~\ref{s:reslock} -- \ref{s:travstand}. In \se{s:reslock} we consider the case of resonance locks created by standing waves. We analyze the resulting tidal efficiency and energetics in \se{s:energ}. In \se{s:nostand} we analyze whether standing waves are able to occur, considering wave breaking in \se{s:wbreak} and critical layers in \se{s:reslockval}. In \se{s:trav} we turn our attention to traveling waves, discussing wave excitation and interference in \se{s:excite} and showing that traveling waves can also create resonance locks in \se{s:trav_intro}. In \se{s:travstand} we then employ numerical simulations to combine our standing and traveling wave results. In \se{s:discuss} we compare our results to observational constraints and discuss physical effects that need to be considered in future work. We then conclude in \se{s:conc} with a summary of our salient results.

\section{Background}\label{s:prelim}
\begin{figure}
  \includegraphics{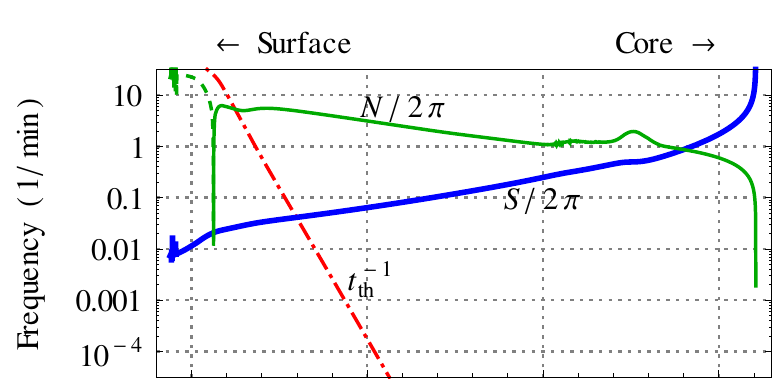}
  \rotatebox{90}{\rule{8pt}{0pt}\rule{0pt}{7pt}$0.2\Msun,\ \Teff = 9,900$ K}\\
  \includegraphics{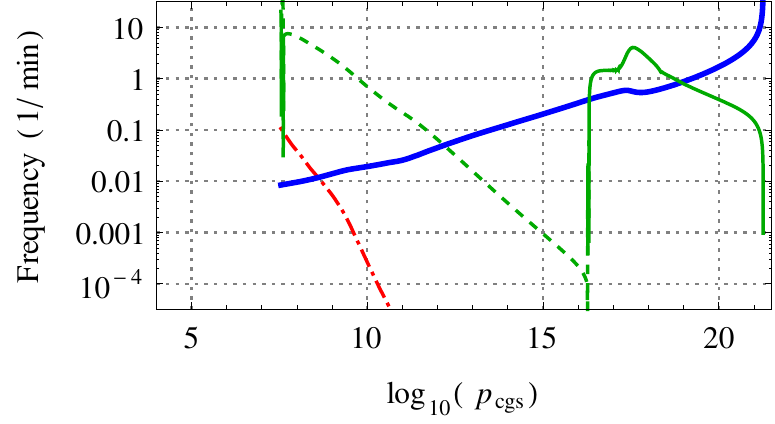}
  \rotatebox{90}{\rule{41pt}{0pt}\rule{0pt}{7pt}$0.2\Msun,\ \Teff = 5,100$ K}\\
  \includegraphics{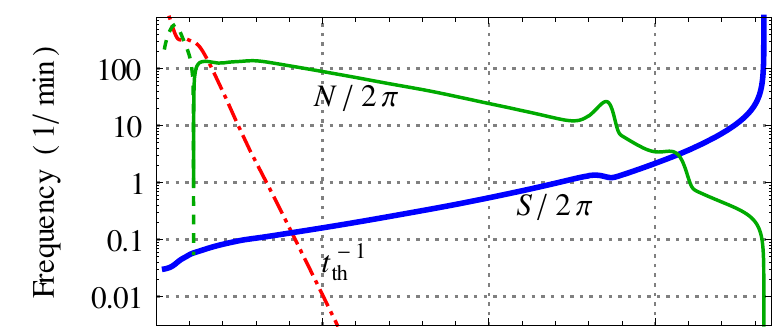}
  \rotatebox{90}{\rule{8pt}{0pt}\rule{0pt}{7pt}$0.6\Msun,\ \Teff = 12,000$ K}\\
  \includegraphics{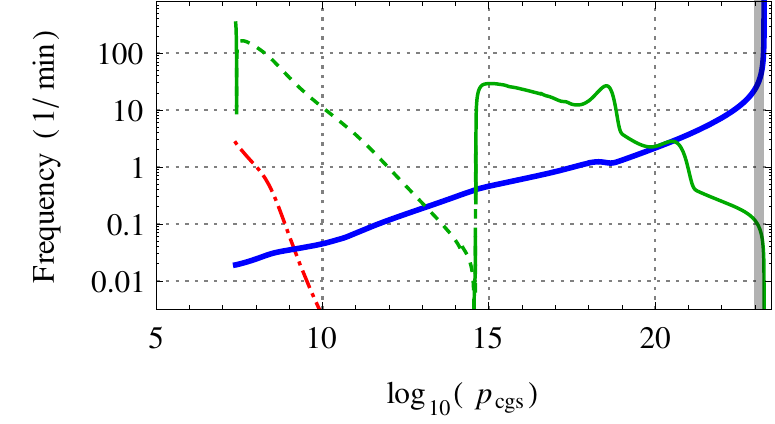}
  \rotatebox{90}{\rule{41pt}{0pt}\rule{0pt}{7pt}$0.6\Msun,\ \Teff = 5,500$ K}\\
  \caption{Propagation diagrams for several of our \wdw{} models listed in \ta{t:wdmods}. The top two panels are our $0.2 \Msun$ \heid{10} and \heid{5} \hewd{}s, with $\Teff=9,900$ and 5,100 K respectively; the bottom two panels are our $0.6\Msun$ \coid{12} and \coid{6} \cowd{}s, with $\Teff=12,000$ and 5,500 K respectively. Each plot shows the \brunt{} frequency $N$ (green line; dashed indicates $N^2<0$), the quadrupolar Lamb frequency $S$ (thick blue line), and inverse local thermal time $1/\ttherm=gF/pc_pT$ (red dot-dashed line). A g-mode is able to propagate where its frequency is less than both $N$ and $S$. In the bottom panel showing our \coid{6} model, the shaded region at high pressure indicates the plasma interaction parameter $\Gamma > 220$, implying crystallization occurs (which is not included in our model); see \se{s:crystal}.}
  \label{f:propdiag}
\end{figure}

\begin{table*}
 \caption{\Wdw{} models. Masses of the helium models (top three) are $0.2\Msun$; masses of the \co{} models (bottom two) are $0.6\Msun$. Helium models each have a hydrogen layer of mass $0.0033M$, and were generated with MESA \citep{paxton11}; \co{} models each have a helium layer of mass $0.017M$ and a hydrogen layer of mass $0.0017M$. Further details on white dwarf models are given in \app{a:models}; \fig{f:propdiag} provides propagation diagrams for several models. The dynamical time is $\tdyn^2=R^3/GM$; the thermal time at the radiative-convective boundary (RCB) is $\ttherm|_\mr{rcb}=pc_p T/gF|_\mr{rcb}$, where we take $2\pi/N=100$ min to define the RCB; the \wdw{} cooling time is $\tcool=\Eth/L$, where $E_\mr{th}=\int c_p T dM$ approximates the total thermal energy; $M_\mr{conv}$ is the mass in the outer convection zone, which increases in size by many orders of magnitude as a \wdw{} cools (see \fig{f:propdiag}); the plasma interaction parameter at the center of the \wdw{} is $\Gamma_\mr{core}=Z^2e^2/kT d_\mr{i}|_{r=0}$, where $Ze$ is the mean ion charge, $d_\mr{i}$ is the ion spacing, and the value of $\Gamma$ corresponding to the onset of crystallization is discussed in \se{s:crystal}; the relativity parameter is $\betas^2=GM/Rc^2$; and $\Is$ is the \wdw{} moment of inertia.}\label{t:wdmods}
\vspace{-1.em}
\begin{center}
\renewcommand{\arraystretch}{1.4}
  \begin{tabular}{lrll.l.l....}\hline
 ID & $\Teff$ (K) & \multicolumn{1}{c}{$L/L_\sun$} & \multicolumn{1}{c}{$R/R_\sun$} & \multicolumn{1}{c}{$\tdyn$ (sec)} & $\ttherm|_{\mr{rcb}}$ (yr) & \multicolumn{1}{c}{$\tcool$ (Gyr)} & \multicolumn{1}{c}{$M_\mr{conv}/M$} & \multicolumn{1}{c}{$\Gamma_\mr{core}$} & \multicolumn{1}{c}{$\betas/10^{-2}$} & \multicolumn{1}{c}{$\Is/MR^2$} \\\hline

 \heid{10} & 9,900 & $1.3\times 10^{-2}$ & 0.038 & 26. & $7.7\times 10^{-8}$ & 0.36 & $3\times 10^{-14}$ & 1.2 & 0.34 & 0.085 \\
 \heid7 & 7,000 & $1.9\times 10^{-3}$ & 0.029 & 18. & $5.1\times 10^1$ & 0.98 & $1\times 10^{-7}$ & 2.7 & 0.38 & 0.11 \\
 \heid5 & 5,100 & $3.9\times 10^{-4}$ & 0.025 & 14. & $1.3\times 10^6$ & 2.5 & $2\times 10^{-4}$ & 5.1 & 0.41 & 0.14 \\
 \hline\coid{12} & 12,000 & $3.0\times 10^{-3}$ & 0.013 & 3.1 & $6.5\times 10^{-9}$ & 0.59 & $1\times 10^{-16}$ & 71. & 0.99 & 0.16 \\
 \coid6 & 5,500 & $1.3\times 10^{-4}$ & 0.013 & 2.9 & $7.3\times 10^2$ & 4.5 & $3\times 10^{-8}$ & 260. & 1.0 & 0.18

\\\hline
\end{tabular}
\end{center}
\end{table*}

A short-period compact object binary efficiently emits gravitational waves that carry off energy and angular momentum. This process causes its orbit to circularize; as such, we will restrict our attention to circular \wdw{} binaries in this work (however, see \citealt{thompson11}). Gravitational waves also cause such a binary's orbit to decay according to $\Omegd=\Omega/\tgw$, where the characteristic gravitational wave inspiral timescale for a circular orbit is \citep{peters64}
\begin{equation}\label{e:tgw}
\tgw = \omegdyn^{-1}\frac{5}{96}\frac{\left( 1+M'/M \right)^{1/3}}{M'/M}\betas^{-5}\left( \frac{\omegdyn}{\Omega} \right)^{8/3}.
\end{equation}
Here $\omegdyn^2=GM/R^3$ is the dynamical frequency of the primary, $M'$ is the mass of the companion, $\betas^2=GM/Rc^2$ is the relativity parameter of the primary, and $\Omega$ is the orbital frequency. The time a binary will take until it begins to transfer mass is given by $\tinsp=3\tgw/8$. For a $0.6 \Msun$ \wdw{} with an equal-mass companion, orbital periods of less than $\sim530$ min imply the binary will begin mass transfer within $10$ Gyr; this restriction reduces to $\Porb\lesssim 270$ min for a $0.2 \Msun$ \wdw{} with an equal-mass companion.

During the process of inspiral, the tidal force acting on each element of the binary steadily grows. The tidal response on a star is typically divided conceptually into two components: the equilibrium tide and the dynamical tide. The equilibrium tide represents the large-scale distortion of a star by a companion's tidal force \citep{zahn77}; it is often theoretically modeled as the filling of an equipotential surface, but can also be treated as the collective nonresonant response of all of a star's eigenmodes. The two viewpoints are equivalent, as in both the tidal forcing frequency is set to zero. Except near very strong resonances, the equilibrium tide contains the great majority of the tidal energy. Nonetheless, whether it produces a strong torque is also influenced by the degree to which it lags behind the tidal potential, which is determined by how strongly the equilibrium tide is damped. For \wdws{}, \citet{willems10} showed that  turbulent convection acting on the equilibrium tide does not cause significant synchronization (\app{a:torque}).

The dynamical tide, on the other hand, corresponds to the tidal excitation of internal stellar waves \citep{zahn75}. In particular, given that tidal forcing periods are much longer than the stellar dynamical timescale, buoyancy-supported gravity waves or g-modes are predominantly excited (although rotationally supported modes become important when the rotation and tidal forcing frequencies become comparable; see \se{s:nonlinrot}). Propagation of gravity waves is primarily determined by the \brunt{} frequency $N$, given by
\begin{equation}\label{e:N} 
 N^2=\frac{1}{g}\left( \frac{1}{\Gamma_1}\frac{d\ln p}{dr} - \frac{d\ln\rho}{dr} \right),
\end{equation}
where $\Gamma_1$ is the adiabatic index. A g-mode is able to propagate where its frequency lies below both $N$ as well as the Lamb frequency $S_l^2 = l(l+1)c_\mr{s}^2/r^2$. Plots of both $N$ and $S_l$ for several helium and \cowd{} models are provided in \fig{f:propdiag}.

Degeneracy pressure satisfies $p\propto\rho^{\Gamma_1}$; substituting this into \eq{e:N} yields $N=0$. Thus the \brunt{} frequency becomes very small in the \wdw{} core where degeneracy pressure dominates gas pressure, scaling as $N^2\propto kT/E_\mr{F}$, where $E_\mr{F}$ is the Fermi energy. Moreover, \wdws{} also often possess outer convection zones with $N^2<0$. As a result, a typical tidally excited g-mode in a \wdw{} possesses both an inner turning point near the core as well as an outer turning point near the radiative-convective boundary.

Lastly, temporarily ignoring degeneracy pressure and assuming an ideal gas equation of state, \eq{e:N} can be expressed as \citep{hansen04}
\begin{equation}
 N^2=\frac{\delad \rho^2 g^2 c_p T}{p^2}\left( \delad -\nabla \right) - g\frac{d\ln\mu}{dr},
\end{equation}
where $\mu$ is the mean molecular weight. From this expression we can see that the \brunt{} frequency becomes larger in composition gradient zones, where $\mu$ decreases with radius. This can be seen in \fig{f:propdiag}, where a ``bump'' in $N$ occurs in helium models due to the helium to hydrogen transition; two bumps are present in \co{} models, resulting from \co{} $\rightarrow$ helium and helium $\rightarrow$ hydrogen.

\section{Dynamical tide regimes in white dwarfs}\label{s:dynregimes}
Here we will give a general overview of the results covered in \S\S~\ref{s:reslock} -- \ref{s:travstand} by enumerating four essential regimes of the dynamical tide in \wdws{}, which comprises the wavelike tidal response. The two basic distinctions made by our four regimes are a) whether tidally excited gravity waves can reflect and become large-amplitude standing modes, or whether they instead represent traveling waves; and b) whether or not a resonance lock can be created. Resonance locks are described in detail in \S\S~\ref{s:reslock} \& \ref{s:trav_intro}; they occur when the tidal torque causes the tidal forcing frequency $\omegt=2(\Omega-\Omegsp)$ to remain constant even as the orbit shrinks.

\newcommand{\SIDnp}[1]{S#1}
\newcommand{\TIDnp}[1]{T#1}
\newcommand{\SID}[1]{(\SIDnp#1)}
\newcommand{\TID}[1]{(\TIDnp#1)}

First, the two regimes where standing waves exclusively occur are:
\begin{enumerate}
 \item[\SIDnp1)] In this regime the dynamical tide represents a purely standing wave, but with a resulting torque that is insufficient to effect a resonance lock even during a perfect resonance. This occurs for long orbital periods or small companion masses. As such, the system quickly sweeps through resonances, and the time-averaged torque is dominated by its value away from resonances. This nonresonant torque is proportional to the eigenmode damping rate, and is thus very small for \wdws{}, due to their long thermal times. As a consequence the average tidal quality factor is very large, tidal heating is negligible, and the spin rate remains essentially constant.

 \item[\SIDnp2)] Here the dynamical tide is again a standing wave, but with eigenmode amplitudes large enough to create resonance locks. This regime is addressed in detail in \se{s:reslock}; we estimate the orbital period corresponding to its onset in \eq{e:Prl}. During a resonance lock, tides become efficient: due to strong tidal torques, the spin frequency changes at the same rate as the orbit decays due to gravitation wave emission. Definite predictions result for the tidal energy deposition rate (\eqp{e:Edot}) and tidal quality factor (\eqp{e:Q}).
\end{enumerate}

Next, the regimes strongly influenced by traveling waves are:
\begin{enumerate}
 \item[\TIDnp1)] In this regime, the off-resonance dynamical tide is still a standing wave, but near resonances the wave amplitude becomes so large that reflection near the surface cannot occur due to wave breaking (\se{s:wbreak}). Furthermore, the traveling wave torque is too weak to cause a resonance lock. Since the typical torque experienced by the \wdw{} is once again the off-resonance standing wave value, the synchronization and tidal heating scenario is very similar to regime \SID1---in other words, tides are ineffective.

 \item[\TIDnp2)] Just as with regime \TID1, the standing wave torque is capped at resonances, becoming a traveling wave; however in this regime the traveling wave torque itself is strong enough to create a resonance lock (terminology discussed further in footnote \ref{fn:resterm}), as addressed in \se{s:trav}. We estimate the onset of this regime in \eq{e:Ptrl}. The predictions for the tidal energy deposition rate and quality factor are the same as in \SID2.
\end{enumerate}

Although we consider only these four regimes in this work, at shorter orbital periods and nearly synchronous rotation, physical effects such as Coriolis modification of stellar eigenmodes and nonlinear tidal excitation mechanisms are likely to become very important; see \se{s:nonlinrot}.

The archetypal scenario is as follows. A \wdw{} binary with a sufficiently long orbital period begins in regime \SID1. Eventually, as the orbit shrinks due to gravitational radiation and the tidal force correspondingly increases in magnitude, the dynamical tide becomes strong enough that a resonance lock takes effect and regime \SID2 is reached. However, as inspiral accelerates, the torque necessary to maintain the resonance lock becomes larger, requiring larger wave amplitudes. When the amplitude becomes too great, the wave begins to break near the outer turning point, and the system enters regime \TID1. Finally, when the traveling wave torque becomes large enough to create a resonance lock, it enters regime \TID2. In \se{s:travstand}, we verify this picture numerically.

\section{Resonance locks by standing waves}\label{s:reslock}
We assume in this section that the dynamical tide is a superposition of standing waves and proceed to predict the tidal evolution of a \wdw{} binary. We discuss the applicability of the standing wave limit in \se{s:nostand}.

Assuming a circular orbit and alignment of spin and orbital angular momenta, the secular tidal torque on a star can be expressed as a sum over quadrupolar ($l=2$) eigenmodes indexed by their number of radial nodes $n$ (\app{a:torque}):
\begin{equation}\label{e:torque1} 
\torq = 8m\Es\epsilon^2 W^2 \sum_nQ_n^2\left[\frac{\omega_n^2\omegt\gamma_n}{(\omega_n^2-\omegt^2)^2+4\omegt^2\gamma_n^2}\right].
\end{equation} 
Here $\omegt = m(\Omega - \Omegsp)$ is the tidal driving frequency in the corotating frame, $\Omega$ is the orbital frequency, $\Omegsp$ is the solid-body rotation rate, $m=2$, $\Es=GM^2/R$ is the \wdw{} energy scale, $\epsilon=(M'/M)(R/a)^3$ is the tidal factor, $M'$ is the companion mass, $a$ is the orbital separation, $W^2=3\pi/10$, and $\gamma_n$ is an eigenmode damping rate (\app{a:lindamp}). Our eigenfunction normalization convention is given in \eq{e:norm}; physical quantities such as the torque are of course independent of the choice of normalization.

The linear overlap integral $Q_n$ appearing in \eq{e:torque1} represents the spatial coupling strength of an eigenmode to the tidal potential, and is normalization dependent. Since the tidal potential spatially varies only gradually, $Q_n$ is large for low-order modes, and becomes much smaller for high-order, short-wavelength modes. We describe various methods of computing $Q_n$ in \app{a:linover}.

The factor in brackets in \eq{e:torque1} describes the temporal coupling of an eigenmode to the tidal potential, and becomes very large during resonances, when the tidal driving frequency becomes close to a stellar eigenfrequency. The nonresonant limit of this factor, which corresponds to the equilibrium tide, increases with stronger damping. Paradoxically, however, the torque during a resonance is inversely proportional to the damping rate, since damping limits the maximum energy a resonant mode attains.

We note that by invoking steady-state solutions to the mode amplitude equations, as we have done here, we fail to account for the energy and angular momentum transfer required to bring a mode's amplitude up to the steady-state value. Additionally, the steady-state solution itself may fail to model the behavior of mode amplitudes very close to resonances correctly; we address this in \se{s:rtunnel}.  Correctly accounting for these two considerations would involve simultaneously solving both the mode amplitude equations for all relevant modes as well as the orbital evolution equations, a task we leave to future study.

Continuing, we focus on resonant tidal effects, and consider the case of a particular eigenmode with a frequency close to the tidal driving frequency, i.e.\ $\omega_n\approx\omegt$. We can then make this substitution everywhere in \eq{e:torque1} other than in the detuning frequency $\detun_n=\omega_n-\omegt$ to find
\begin{equation}\label{e:torque2} 
\torq \approx 2m\Es\epsilon^2 W^2 Q_n^2\left(\frac{\omega_n\gamma_n}{\detun_n^2+\gamma_n^2}\right),
\end{equation}
having dropped nonresonant terms. One might expect that since the strongest torques are achieved very near resonance, where $\detun_n\sim 0$, a system should evolve quickly through resonances, and they should have little effect on the long-term orbital and spin evolution. This is often accounted for by using a ``harmonic mean'' of the torque to produce a synchronization time, i.e., $\tsync=\int \Is d\Omega / \torq$, where $\Is$ is the moment of inertia \citep{goodman98}.

Under particular circumstances, however, it is possible to achieve a resonance lock, where an eigenmode remains in a highly resonant state for an extended period of time, as originally proposed by \citet{witte99}. Very near a resonance, the torque depends very strongly on the detuning frequency $\detun_n$, and very weakly on the orbital period by itself; as a result, the essential criterion that must be satisfied for a resonance lock to occur is that the detuning frequency must remain constant:
\begin{equation}\label{e:reslock} 
0=\dot{\detun}_n=m\left((1-C_n)\Omegspd-\Omegd\right),
\end{equation} 
where $\partial\omega_n/\partial\Omegsp=-mC_n$ accounts for rotational modification of the stellar eigenmodes, $C_n\approx 1/6$ for high-order $l=2$ g-modes and slow rotation  \citep{unno89}, and we have assumed the \wdw{} rotates as a solid body. (We justify the solid-body rotation assumption in \se{s:reslockval}.)

For simplicity, we will henceforth ignore rotational modification of the stellar eigenfrequencies, so that $C_n\rightarrow0$. This limits the quantitative applicability of our results to where $\Omegdiff\gtrsim\Omegsp$. We also neglect progressive \wdw{} cooling, which decreases the \brunt{} frequency and consequently lowers eigenmode frequencies; this is valid so long as the cooling time $\tcool$, which is on the order of $\sim$~Gyr for the models listed in \ta{t:wdmods}, is much longer than the gravitational wave decay time $\tgw$ (\eqp{e:tgw}). Subject to these simplifications, \eq{e:reslock} then reduces to $\Omegd=\Omegspd$, i.e., that the orbital and spin frequencies evolve at the same rate. Since the orbital frequency increases due to the emission of gravitational waves, and the spin frequency increases due to tidal synchronization, this phenomenon is plausible at first glance. We now work out the mathematical details.

The evolution of $\Omegsp$ and $\Omega$ proceed as
\begin{equation}\label{e:omegdots} 
\begin{pmatrix}\Omegspd/\Omegsp\\\Omegd/\Omega\end{pmatrix} = \begin{pmatrix}\torq/\Is\Omegsp\\ 1/\tgw+(3/2)(\Edtide/|\Eorb|) \end{pmatrix},
\end{equation} 
where $\Edtide$ is the secular tidal energy transfer rate\footnote{Our convention is that $\torq>0$ or $\Edtide>0$ implies that orbital angular momentum or energy is being transfered to the \wdw{}(s).} and the gravitational wave inspiral time $\tgw$ is given in \eq{e:tgw}. Here we have failed to account for tidal effects in the companion, which would provide an extra contribution to $\Edtide$; see below.

Using equations \eqref{e:omegdots} and \eqref{e:etau}, \eq{e:reslock} becomes
\begin{equation}
\frac{\Omega}{\tgw} = \torq\left( \frac{1}{\Is} - \frac{3}{2}\frac{1}{\mu a^2} \right),
\end{equation}
where $\mu=MM'/(M+M')$ is the reduced mass. Since $\mu a^2\gg \Is$, in the present context we can neglect any tidal influence on $\Omegd$ even in this extreme-resonance scenario; this also now justifies dropping the companion's contribution to $\Edtide$. We can then approximate the resonance lock criterion as
\begin{equation}\label{e:reslock2} 
\frac{\Omega}{\tgw} = \frac{\torq}{\Is}.
\end{equation}
This implies that in a resonance lock, the torque increases as $\tau\propto\Omega^{11/3}$ as the orbit decays.

\begin{figure}
  \begin{overpic}{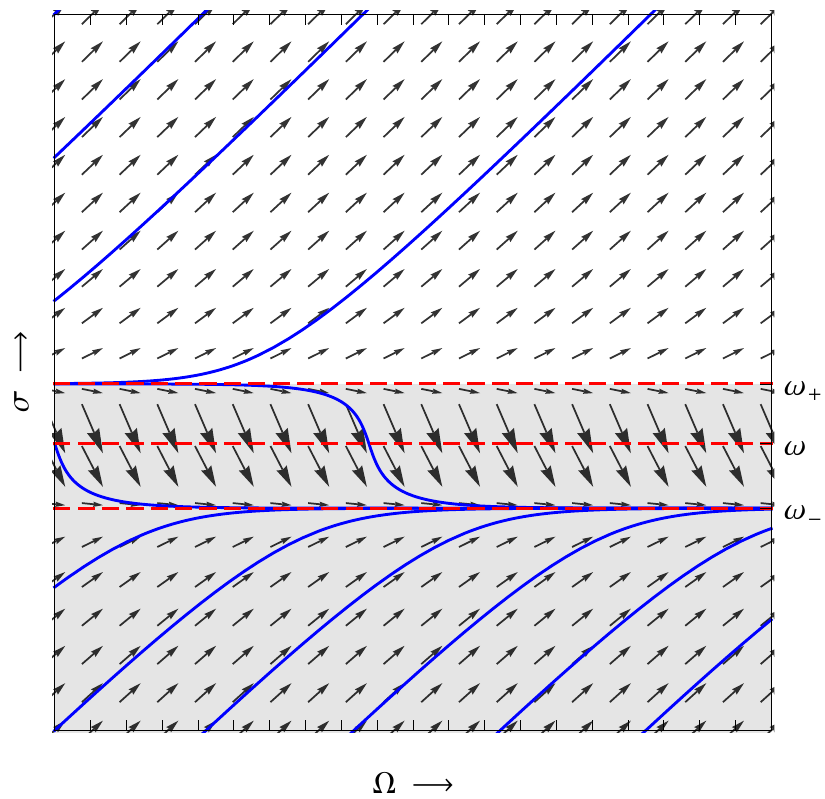}
  \end{overpic}
  \caption{Plot schematically illustrating the dynamics of a resonance lock. The abscissa is the orbital frequency $\Omega$, which increases due to gravitational wave radiation, and the ordinate is the Doppler-shifted $l=m=2$ tidal driving frequency $\omegt=2(\Omegdiff)$. The arrows depict the vector field describing the orbital evolution equations (\eqp{e:omegdots}). The eigenfrequency of the included mode is $\omega$, which is flanked by stable and unstable fixed points of the evolution equations: $\omega_-$ and $\omega_+$, respectively. The dashed red horizontal lines show these three frequencies, while the blue curves are example system trajectories. The stable point $\omega_-$ corresponds to a resonance lock, while the unstable fixed point $\omega_+$ corresponds to the upper boundary of the stable fixed point's basin of attraction (shaded region). In producing this plot, the correct functional form of the equation of motion has been used, except with $\tgw$ and $\epsilon$ taken as constant, and with artificially chosen values of the various germane parameters. In particular, for realistic resonance lock situations in \wdws{}, the three frequencies shown are very close together, and the basin of attraction barely extends beyond $\omega_-$.}
  \label{f:vecplot}
\end{figure}

Once \eq{e:reslock2} is satisfied, the lock can persist for a long period of time, since the quantities in \eq{e:torque2} affecting the magnitude of the torque other than $\delta\omega_n$ change only very gradually in time. (We verify this using numerical orbital evolution simulations in \se{s:travstand}.) In other words, a resonance lock represents a dynamical attractor; \fig{f:vecplot} provides an illustration of the nonlinear dynamics behind a resonance lock. The lock could eventually be destroyed if the mode responsible began to break near its outer turning point, which would drastically increase the effective damping rate. This phenomenon is explained in \se{s:nostand}.

If we assume that the system begins completely unsynchronized, so that $\omegt=2\Omega$, we can determine the orbital period where a lock first occurs, which we denote $\Prl$, by substituting equations \eqref{e:torque2} and \eqref{e:tgw} into \eqref{e:reslock2}, setting $\omega_n=\omegt$ and hence $\detun_n=0$, and then solving for the orbital period. To this end, we invoke the following approximate scalings for the eigenmode linear tidal overlap integral $Q$ and damping rate $\gamma$:
\begin{displaymath}
 Q\approx Q_0 (\omegt/\omegdyn)^a\quad\text{and}\quad\gamma\approx\gamma_0(\omegt/\omegdyn)^{-b},
\end{displaymath}
where values of the various parameters in these expressions are listed in \ta{t:eigprops} for our fiducial \wdw{} models. Scaling parameter values to those for our \coid{6} model, we have
\begin{equation}\label{e:Prl}
\Prl \sim 170\ \mr{ min}\left( \frac{\tdyn}{2.9\ \mr{s}} \right)F_\rl^p,
\end{equation}
where $\tdyn=(R^3/GM)^{1/2}$ is the \wdw{}'s dynamical time, the factor $F_\rl$ is
\begin{equation}\begin{split}\label{e:Frl} 
F_\rl&\sim\left( \frac{M'}{M} \right)\left( \frac{1+M'/M}{2} \right)^{-5/3}\\
&\times\left( \frac{\betas}{0.010} \right)^{-5}\left( \frac{\Is}{0.18 MR^2} \right)^{-1} \\
&\times\left( \frac{Q_0}{27} \right)^{2}\left( \frac{\gamma_0}{2.9\times10^{-14}\omegdyn} \right)^{-1}
\\&\times \left(2.15\times10^{26}\right)(0.00327)^{1/p},
\end{split}\end{equation}
the power $p$ is in general
\begin{equation}
 p=\frac{1}{4/3+2a + b}\ll1,
\end{equation}
and $p=0.094$ for our \coid{6} model (\ta{t:eigprops}). (The last line of \eqp{e:Frl} is equal to unity for $p=0.094$.)

For comparison, direct numerical evaluation of eigenmode properties with our \coid{6} \wdw{} model yields $\Prl = 170$ min for an equal-mass companion, due to an $n=122$ g-mode. This is in very good agreement with the analytic approximation in \eq{e:Prl}, and is also consistent with our numerical results in \se{s:travstand}. We provide values of $\Prl$ for each of our fiducial models in \ta{t:tideprops}. These results show that resonance locks begin at longer orbital periods for cooler \wdw{} models, due to larger outer convective zones and longer overall diffusive damping times (smaller $\gamma_{n}$; see \ta{t:eigprops} \& \fig{f:gammaplots}). This increases the maximum possible tidal torque, which is proportional to $1/\gamma_{n}$ (\eqp{e:torque2}).

\begin{table}
 \caption{\Wdw{} tidal parameters. For each of our fiducial \wdw{} models from \ta{t:wdmods}, we list the orbital period $\Prl$ of its first resonance lock (\se{s:reslock}, \eqp{e:Prl}), the orbital period $\Ptrl$ below which traveling wave resonance locks can occur (\se{s:effres}, \eqp{e:Ptrl}), its value of $\lambda$ (\se{s:energ}, \eqp{e:tideE}), and the tidal quality factor $\Qt|_{100\ \mr{min}}$ for a resonance lock (\se{s:Q}, \eqp{e:Q}) evaluated at $\Porb=100$ min. All values are for an equal-mass companion. We determined $\Prl$ and $\Ptrl$ by directly searching over numerically computed eigenmode properties.}\label{t:tideprops}
\vspace{-1.em}
\begin{center}
\renewcommand{\arraystretch}{1.4}
  \begin{tabular}{lrc.c}\hline
 ID & \multicolumn{1}{c}{$\Prl$ (min)} & \multicolumn{1}{c}{$\Ptrl$ (min)} & \multicolumn{1}{c}{$\lambda/10^{-2}$} & $\Qt|_{100\ \mr{min}}$\\\hline
{\tt He10} & 67 & 49 & 1.2 & $2\times 10^9$\\
{\tt He7} & 270 & 49 & 2.4 & $1\times 10^9$\\
{\tt He5} & 1,400 & 90 & 4.6 & $1\times 10^9$\\
\hline{\tt CO12} & 31 & 22 & 6.3 & $1\times 10^7$\\
{\tt CO6} & 170 & 40 & 7.4 & $1\times 10^7$

\\\hline
\end{tabular}
\end{center}
\end{table}

\section{Energetics}\label{s:energ}
\subsection{Tidal quality factor}\label{s:Q} 
A star's tidal quality factor $\Qt$ can be defined as
\begin{equation}\label{e:Qdef} 
\Qt = \frac{\Omega \Etide}{\Edtide},
\end{equation} 
where the energy content of the tide $\Etide$ is approximately given by (\app{a:torque})
\begin{equation}\label{e:tideE} 
\Etide = \lambda \epsilon^2 \Es,
\end{equation} 
$\lambda=2W^2\sum_nQ_n^2$, and values of $\lambda$ for various \wdw{} models are given in \ta{t:tideprops}. Using the relationship between the tidal torque and energy transfer rate from \eq{e:etau}, we see that the tidal torque can be expressed in terms of $\Qt$ by
\begin{equation}\label{e:Qtorq} 
 \torq = \frac{\Etide}{\Qt}.
\end{equation} 
Since $\Etide\sim \lambda F_\mr{tide}\,h$, where $h\sim\epsilon R$ is the height of the equilibrium tide and $F_\mr{tide}\sim\epsilon\Es /R$ is the tidal force, we see that our definition of $\Qt$ is consistent with $1/\Qt$ representing an effective tidal lag angle; see e.g.\ \citet{goldreich66}. Note that since $\Qt$ parameterizes the total tidal energy deposition rate, which includes mechanical energy transfer associated with increasing the \wdw{} spin, the value of $\Qt$ alone does not fully determine the tidal heating rate; see \se{s:heat}.

Using the previous three equations along with the resonance lock condition from \eq{e:reslock2}, we have that the value of $\Qt$ during a resonance lock is
\begin{equation}\label{e:Q0}
\Qt = \frac{\lambda\epsilon^2 \tgw \Es}{\Is\Omega}.
\end{equation}
Substituting further yields
\begin{equation}\begin{split}\label{e:Q} 
\Qt &\approx 9.7\times 10^6 \left(\frac{\Porb}{100\ \mr{min}}\right)^{-1/3} \left( \frac{\tdyn}{2.9\ \mr{s}} \right)^{1/3}\\
&\times\left( \frac{M'}{M} \right)\left( \frac{1+M'/M}{2} \right)^{-5/3}\left( \frac{\lambda}{0.074} \right)\\
&\times\left( \frac{\Is}{0.18 MR^2} \right)^{-1}\left( \frac{\betas}{0.010} \right)^{-5}.
\end{split}\end{equation}
Here $\Is$ is the moment of inertia, $\betas^2=GM/Rc^2$, and all values have been scaled to those appropriate for our \coid{6} model (Tables \ref{t:wdmods} \& \ref{t:tideprops}).

\Eq{e:Q} is a central result of this paper. It is independent of eigenmode properties and is only weakly dependent on the orbital period, although it depends strongly on the mass, radius, and companion mass. Eigenmode properties do of course dictate when this value of $\Qt$ is applicable, i.e., when resonance locks are able to occur. We will further show in \se{s:trav_intro} that \eq{e:Q} can hold even when the dynamical tide is a traveling wave, and the standing wave formalism presented thus far is invalid.

Values of the various quantities entering into \eq{e:Q} are provided for a selection of helium and \cowd{} models in \ta{t:wdmods}. In particular, since the inspiral time is much longer for low-mass \hewd{}s than for more massive \cowd{}s, \eq{e:Q} predicts that the tidal quality factor $\Qt$ should be much larger ($\sim100\times$) for \hewd{}s, as shown in \ta{t:tideprops}, meaning tidal effects are more efficient in \cowd{}s.

\subsection{Tidal heating}\label{s:heat} 
\newcommand{\Pasync}{\delta P}
The rate at which heat is dissipated in the \wdw{} assuming solid-body rotation can be derived using \eq{e:etau}:
\begin{equation}
\begin{split}
  \Edheat &= \Edtide - \dot{E}_\mr{mech}\\
  &= \Omega\torq - \frac{d}{dt}\left(\frac{1}{2}\Is \Omegsp^2\right)\\
  &=\Is\Omegspd\dOmeg = \frac{\Etide\,\dOmeg}{\Qt},
\end{split}
\end{equation} 
where $\dOmeg=\Omega-\Omegsp$. During a resonance lock we have $\Omegspd\approx\Omegd=\Omega/\tgw$, so that
\begin{equation}
\Edheat \approx \frac{\Is\Omega\;\dOmeg}{\tgw},
\end{equation} 
with $\dOmeg$ then being approximately constant (having neglected rotational modification of \wdw{} eigenmodes; see \se{s:nonlinrot}). Defining the asynchronicity period as $\Pasync=2\pi/\dOmeg$, we can evaluate this further as
\begin{equation}\begin{split}\label{e:Edot} 
\Edheat &\approx 1.4\times10^{-2}\;L_\sun\left( \frac{M'}{M} \right)\left( \frac{1+M'/M}{2} \right)^{-1/3}\\
&\!\!\!\!\!\!\times\left(\frac{\Porb}{10\ \mr{min}}\right)^{-11/3}\!
\left(\frac{\Pasync}{200\ \mr{min}}\right)^{-1}\\
&\!\!\!\!\!\!\times\left( \frac{\Is}{0.18 MR^2} \right)\left(
\frac{M}{0.6\Msun} \right)^{8/3}\!\left( \frac{R}{0.013R_\sun} \right)^{2},
\end{split}\end{equation}
again scaling variables to our \coid{6} model's properties (\ta{t:wdmods}).

As a simple analytical estimate, consider the example of a resonance lock beginning with the \wdw{} unsynchronized at an orbital period $P_0$ and continuing until the Roche period of $P_\mr{Roche}\sim \tdyn \ll P_0$. The total orbital energy dissipated in the \wdw{} as heat in this example is
\begin{equation}
\begin{split}
\Delta E_\mr{heat} &= \frac{2\pi \Is}{P_0} \int \frac{\Omega}{\tgw}dt\\
  &\approx \frac{4\pi^2 \Is}{\tdyn P_0}\\
  &\sim 7.0\Es \left( \frac{\tdyn}{P_0} \right)\left( \frac{\Is}{0.18MR^2} \right),
\end{split}
\end{equation} 
which could be very large depending on the value of $P_0$. If for $P_0$ we use our estimate from \se{s:reslock} of $\Prl\sim 170$ min appropriate for our \coid{6} model, we have $\Delta E_\mr{heat} \sim 2\times10^{47}$~ergs, a factor of $\sim 3$ larger than the \coid{6} model's thermal energy.

Tidal heating can directly add to a \wdw{}'s luminosity and minimally affect its thermal structure if a) the thermal time $\ttherm=pc_pT/gF$ at the outer turning point, where wave damping is most efficient, is smaller than $\tgw$, and b) $\Edheat\ll L$. The outer turning point occurs due to the outer convection zone, so $\ttherm|_\mr{rcb}$ (radiative-convective boundary) is an appropriate value to use. We find $\ttherm|_\mr{rcb}\lesssim10^6$ years for all of our \wdw{} models, as shown in \ta{t:wdmods}; this is $\ll\tgw$ for $\Porb\gtrsim10$ min, implying that criterion (a) is satisfied. Moreover, for all models other than our $\Teff=5,100$~K helium and $\Teff=5,500$~K \cowd{}s, $\ttherm|_\mr{rcb}\lesssim50$ years, and $\ttherm|_\mr{rcb} \lesssim \tgw$ is satisfied even directly prior to mass transfer.

Criterion (b) above is more restrictive: examining \eq{e:Edot} shows that near orbital periods of $\sim10$ min, the tidal heating rate approaches typical \wdw{} luminosities. The orbital period where this occurs depends weakly on the various parameters appearing in \eq{e:Edot}, since $\Edheat\propto\Porb^{-11/3}$. Thus the thermal structure of \wdw{}s in close binaries may adjust significantly to accommodate the additional heat input for $\Porb\lesssim 10$ min. We discuss the consequences of tidal heating further in \se{s:discheat}.

\subsection{Tidally enhanced orbital decay} \label{s:Pdot}
\newcommand{\Pd}{\dot{P}}
\newcommand{\Pdotgw}{\Pd_\mr{gw}}
\newcommand{\Porbd}{\Pd_\mr{orb}}
\newcommand{\Pdottide}{\Pd_\mr{tide}}
Although the rate $\Porbd$ at which the orbital period of an inspiraling \wdw{} binary decays is dominated by the gravitational wave term $\Pdotgw=-\Porb/\tgw$, tidal energy dissipation implies a small deviation from this value (see also \citealt{piro11}).\footnote{Note that the purpose of this section is to determine the influence of the tidal energy deposition term $\Edtide$ on the rate of orbital decay, even though this term is neglected everywhere else in this work, as justified in \se{s:reslock}.} We can compute this difference for a system consisting of two \wdw{}s both undergoing resonance locks by using equations \eqref{e:reslock}, \eqref{e:omegdots}, and \eqref{e:etau}, which yield
\begin{equation}
 \Porbd = \Pdotgw + \Pdottide,
\end{equation} 
where
\begin{equation}\label{e:Pdottide}
 \Pdottide = \left( \frac{S}{1-S} \right)\Pdotgw,\quad S \approx 3\left( \frac{I_1+I_2}{\mu a^2} \right),
\end{equation}
$I_{1,2}$ are the moments of inertia of the two \wdw{}s, $\mu=M_1 M_2/(M_1+M_2)$ is the reduced mass,  $a$ is the semi-major axis, and we have again neglected rotational modification of \wdw{} eigenmodes (i.e.\ $\partial\omega_n/\partial\Omegsp=0$, where $\omega_n$ is a corotating-frame eigenfrequency; see \se{s:nonlinrot}). This effect may be detectable in future observations of close \wdw{} binaries, as discussed in \se{s:discobs}.

\section{Applicability of standing waves}\label{s:nostand} 

\subsection{Wave breaking}\label{s:wbreak}
\begin{figure*}
  \centering
  \includegraphics{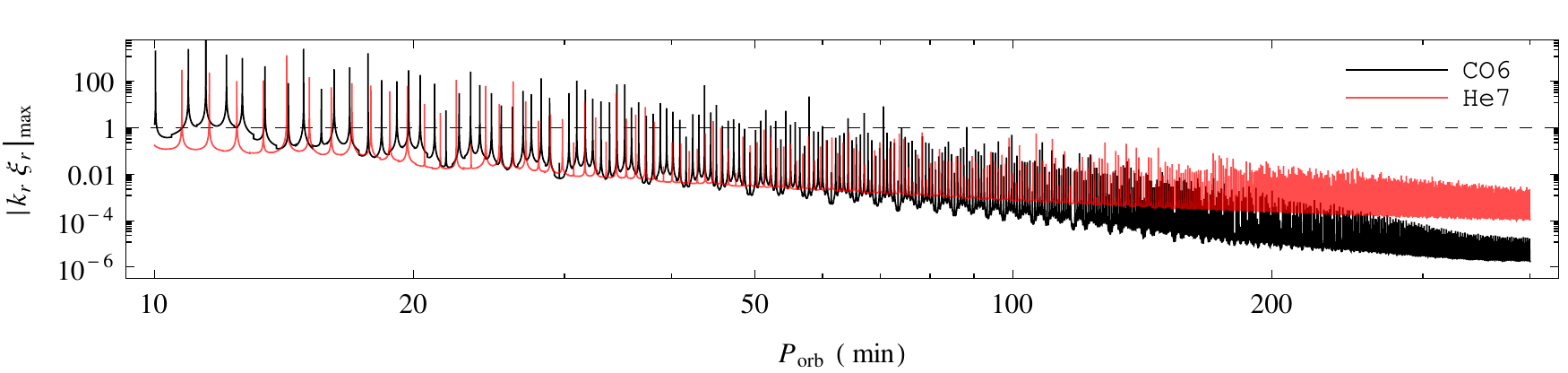}
  \caption{The maximum value of $\krxirabs$ attained throughout the propagation cavity of our \coid{6} (black line) and \texttt{He7} (red line) \wdw{} models (\ta{t:wdmods}), assuming adiabatic standing waves, $\Omegsp = \Omega/2$, and an equal-mass companion. Where $\krxirm>1$, wave breaking occurs (\se{s:wbreak}), and the effective wave damping time becomes roughly the group travel time across the \wdw{}. This occurs very near resonances for $\Porb\lesssim0.5$ -- 1 hr.}
  \label{f:kxplot}
\end{figure*}

It is important to determine whether the dynamical tide we are attempting to study represents a standing wave or a traveling wave. If it is a standing wave, meaning it is able to reflect at its inner and outer turning points without being absorbed, then it can achieve large amplitudes due to resonances with tidal forcing frequencies (as assumed in \se{s:reslock}). In the absence of nonlinear effects that can occur at large amplitudes, a standing wave's damping rate is well approximated by the quasiadiabatic value (\app{a:lindamp}), which is small for \wdws{} due to their high densities and long thermal times (see \fig{f:gammaplots}). On the other hand, if the dynamical tide instead behaves as a traveling wave, resulting from absorption prior to reflection, then its damping time is approximately a group travel time.

In this section we determine whether the nonlinear process of gravity wave breaking causes tidally excited g-modes in \wdws{} to be absorbed near the surface and hence to become traveling waves, as has been suggested in recent studies \citep{fuller11,fuller12}. Gravity wave breaking has been considered extensively in the atmospheric science community, since it occurs in Earth's atmosphere; see e.g.\  \citet{lindzen81}. It is also thought to occur in the cores of solar-type stars \citep{goodman98,barker10}.

Breaking occurs when a wave's amplitude becomes large enough to disrupt the stable background stratification. One way to derive the condition under which this happens is to determine when a wave would produce its own convective instability, which is equivalent to the perturbed \brunt{} frequency (squared) becoming comparable to the background value---this then makes the total value negative, implying convection. The Eulerian perturbation to $N^2$ is given in linear theory by
\begin{equation}
\frac{\delta N^2}{N^2} \approx k_r \xir - \frac{\delta p}{p} - \frac{\xir}{H_\rho} + \frac{\delta g}{g},
\end{equation} 
where $H_\rho$ is the density scale height and $k_r$ is the wavenumber in the direction of gravity. Since $k_r \xir$ is much larger in magnitude than the other terms for g-modes, the wave breaking condition thus becomes
\begin{equation}\label{e:kxcrit} 
\krxirabs\sim 1.
\end{equation} 

Other nonlinear processes also come into play when $\krxirabs\sim1$. Indeed, this criterion is equivalent to $\mr{Ri}\sim 1/4$, where $\mr{Ri}$ is the Richardson number due to the wave's shear, which implies the wave is Kelvin-Helmholtz unstable. \Eq{e:kxcrit} is also similar to the condition under which surface ocean waves break: when the vertical displacement becomes comparable to the wavelength.

To determine whether g-modes break, we evaluated the linear, quadrupolar tidal fluid response assuming global adiabatic normal modes and an equal-mass companion; see \app{a:globlin}. Under these assumptions, we find that for both our helium and \cowd{} models, the dynamical tide breaks for close resonances at orbital periods as large as $\sim1$ hr, as shown in \fig{f:kxplot}. The off-resonance dynamical tide begins to break more generically at $\Porb\lesssim10$~--~20~min.\footnote{This is in conflict with the claims made in \citet{fuller12}, since that work used $k_r|\vecxi|\sim 1$ to assess wave breaking, instead of \eq{e:kxcrit}. The total displacement $|\vecxi|=(\xir^2+\xih^2)^{1/2}$ includes horizontal motion, which is perpendicular to the stratification and thus does not contribute to breaking. As a g-mode's horizontal motion is much greater than its vertical motion, \citet{fuller12} overestimated the degree of breaking by a factor of $\sim\xih / \xir \sim \omegdyn/\omegt \gg 1$, where $\omegt=2(\Omega-\Omegsp)$ is the $l=m=2$ tidal driving frequency.} Furthermore, for all of the \wdw{} models we have considered (\ta{t:wdmods}), we find that at sufficiently long orbital periods, the dynamical tide doesn't break even for a perfect resonance, and thus that standing wave resonance locks should be able to occur. As such, we expect wave breaking not to operate during a significant portion of the inspiral epoch, in which case the analysis presented in \se{s:reslock} may be valid. We address this in more detail in \se{s:resnum}.

\subsection{Differential rotation and critical layers}\label{s:reslockval}
The possibility of differential rotation represents a significant challenge to the standing wave assumption we utilized in \se{s:reslock}. Indeed, tidal angular momentum is preferentially deposited in the outer layers of a \wdw{}, since that is where damping times are shortest and waves are able to communicate their energy and angular momentum content to the background stellar profile \citep{goldreich89b}. Thus tides do not naturally induce solid-body rotation, and instead tend to first synchronize layers near the outer part of the gravity wave propagation cavity \citep{goldreich89}, absent the influence of efficient internal angular momentum transport.

The presence of a synchronized or ``critical'' layer at the edge of a mode propagation cavity implies that the mode's corotating frequency tends to zero at that location, which in turn means its radial wavenumber becomes very large due to the asymptotic g-mode dispersion relation $\omega\sim N (k_{\h}/k_r)$, where  $k_\h$ and $k_r$ are respectively the perpendicular and radial wavenumbers. As a result, the mode's local damping time becomes very short, and it is absorbed rather than reflected, eliminating the possibility of achieving resonant amplitudes (although traveling waves can also effect resonance locks at short orbital periods; \se{s:trav}).

In \app{a:trans} we analyze angular momentum redistribution by fossil magnetic fields, possibly generated by a progenitor star's convective core (during hydrogen or helium fusion) and amplified by flux freezing as the core contracts. We calculate that a field strength of only $\sim200$~G is required to maintain solid-body rotation during a resonance lock for an orbital period of $\sim100$~min in our \coid{6} model, and only $\sim20$~G in our \heid{7} model (\ta{t:wdmods}). \citet{liebert03} conclude that at least $\sim10$\% of \wdws{} have fields $\gtrsim10^6$~G, and speculate that this fraction could be substantially higher; field strengths in \wdw{} interiors may be even more significant. With a field of $10^6$~G, our calculations indicate that critical layers should not occur until orbital periods of less than 1~min, or even less if the field can wind up significantly without becoming unstable.

\subsection{Validity of the secular approximation}\label{s:rtunnel}
The Lorentzian mode amplitude solutions invoked in \app{a:globlin} to produce the standing wave torque in \eq{e:torque1} are strictly valid only when a mode's amplitude changes slowly relative to its damping time. Further examining \eq{e:torque2}, we see that near a perfect resonance the amplitude changes by a factor of $\sim2$ as the detuning frequency changes by of order the damping rate $\gamma_n$. Thus the Lorentzian solution is applicable near a perfect resonance only when
\begin{equation}\label{e:tunnel} 
 \gamma_n^{-1}\lesssim \tgw \frac{\gamma_n}{\Omega},
\end{equation} 
which evaluates to
\begin{equation}\label{e:tunnel2} 
 \Porb \gtrsim 90\ \mr{min} \left( \frac{M}{0.6M\sun} \right)^{5/11} F_\mr{sec},
\end{equation} 
where
\begin{equation*}
 F_\mr{sec} = \left( \frac{M'}{M} \right)^{3/11}\left(\frac{1+M'/M}{2}\right)^{-1/11} \left( \frac{\gamma_n^{-1}}{80\ \mr{yr}} \right)^{6/11}.
\end{equation*} 

\Eq{e:tunnel2} is scaled to values for our \coid6 model (\fig{f:rlockco}); the restriction instead evaluates to $\Porb\gtrsim50$ min for our \heid7 model, using a damping time of $\gamma_n^{-1}\sim60$ yr appropriate for the initial resonance lock. Below these periods, the Lorentzian solution becomes invalid and the exact outcome is unclear, although our preliminary numerical integrations of fully coupled mode amplitude and orbital evolution equations indicate that resonance locks can still occur even beyond the validity of the Lorentzian solution. (We address a similar concern relating to angular momentum transport in \app{e:transirl}.) Nonetheless, we find that the initial standing wave resonance lock occurs at orbital periods larger than the critical value from \eq{e:tunnel2} in our \coid6 and \heid7 models  (\ta{t:tideprops}), meaning resonance locks should proceed as expected.

\section{Traveling waves}\label{s:trav}
\subsection{Excitation and interference}\label{s:excite}
In this section, we will describe two different mechanisms of tidal gravity wave excitation considered in the literature. We will then compare both sets of theoretical predictions to our numerical results to assess which mechanism predominantly operates in our fiducial \wdw{} models.

\begin{figure*}\centering
\begin{tabular}{lr}
  \begin{overpic}{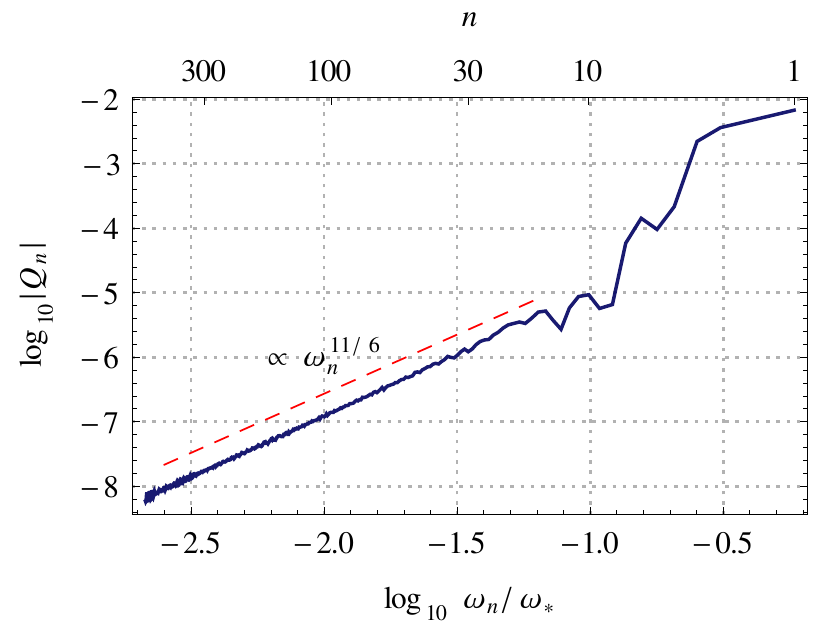}
  \put(55,18){\colorbox{white}{\heid5: $0.2\Msun,\ \Teff = 5,100$ K}}
  \put(20,55){\colorbox{white}{\Large 1}}
  \end{overpic}
&
  \begin{overpic}{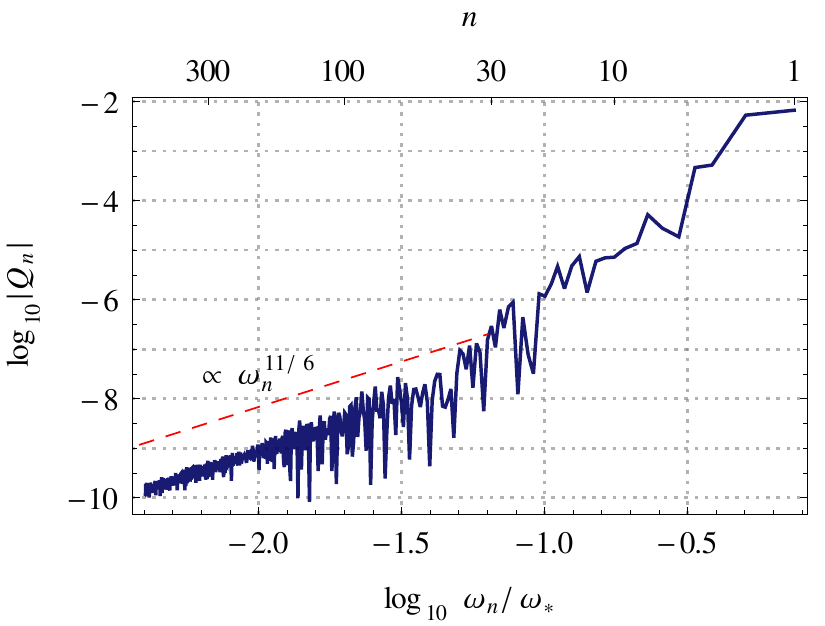}
  \put(55,18){\colorbox{white}{\heid7: $0.2\Msun,\ \Teff =7,000$ K}}
  \put(20,55){\colorbox{white}{\Large 2}}
  \end{overpic}\\
  \begin{overpic}{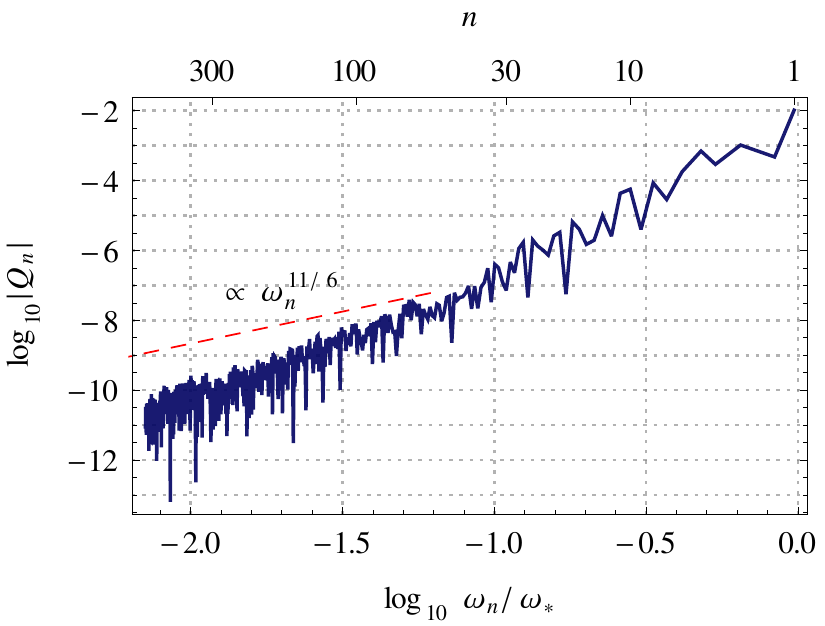}
  \put(52,18){\colorbox{white}{\heid{10}: $0.2\Msun,\ \Teff =9,900$ K}}
  \put(20,55){\colorbox{white}{\Large 3}}
  \end{overpic}
&
  \begin{overpic}{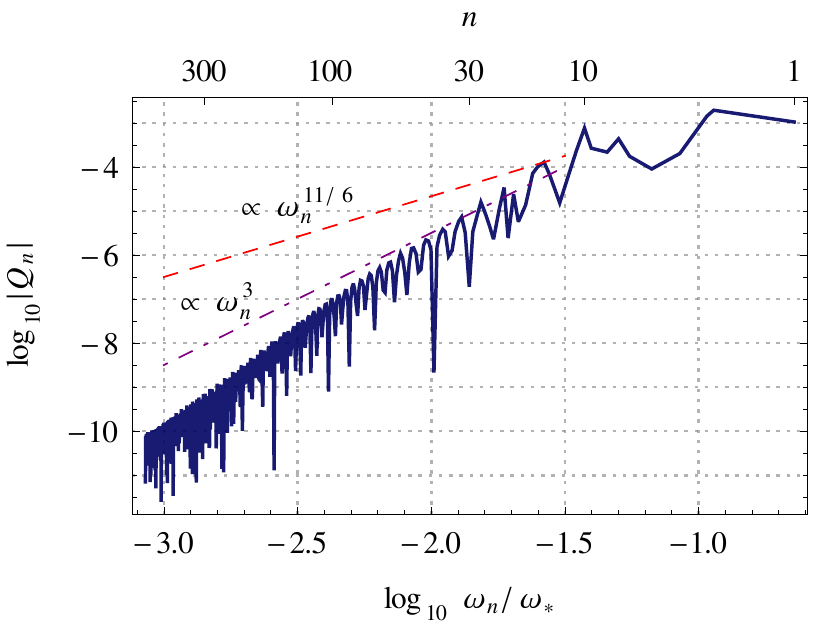}
  \put(55,18){\colorbox{white}{\coid{6}: $0.6\Msun,\ \Teff =5,500$ K}}
  \put(20,55){\colorbox{white}{\Large 4}}
  \end{overpic}
\end{tabular}
  \caption{Plots of the linear tidal overlap integral $Q_n$, which characterizes the spatial coupling strength between the tidal potential and a given mode (\app{a:linover}), as a function of the eigenmode frequency $\omega_n$ and radial order $n$, for the first 500 g-modes in four of our fiducial \wdw{} models (\ta{t:wdmods}). Panels 1 -- 3 are \hewd{}s ordered by increasing temperature; panel 4 is a \cowd{}. A smooth power law scaling of $Q_n\propto\omega_n^{11/6}$ implies that gravity wave excitation by the tidal potential occurs at the interface between a \wdw{}'s outer convection zone and its inner radiative core (\se{s:excite}); this can be seen in the cooler helium models from panels 1 \& 2. Hotter \wdws{} have smaller convective regions, and wave excitation instead may occur at composition gradient zones \citep{fuller12}; this mechanism predicts steeper, more jagged profiles of $Q_n$ with $\omega_n$, as in panels 3 \& 4.}
  \label{f:Qplots}
\end{figure*}

\citet{zahn75} showed that when a gravity wave is well described by its WKB  solution, a conserved wave energy flux results. Thus gravity waves must be excited where the WKB approximation is invalid: where the background stellar model---particularly the \brunt{} frequency $N$---changes rapidly relative to a wavelength.

One natural candidate for wave excitation, then, is at a radiative-convective boundary (RCB), where $N^2$ abruptly becomes negative. \Wdw{}s possess convective envelopes near their surfaces (\fig{f:propdiag}), so this mechanism is plausible. The resulting theoretical prediction \citep{zahn75,goodman98} is that the traveling wave tidal torque should scale as $\torq\propto\omegt^{8/3}$, where $\omegt=2(\Omegdiff)$ is the $m=2$ tidal driving frequency. Using our calculation of the traveling wave torque in \eq{e:travtorque} from \se{s:trav_intro}, we see that this in turn implies that the linear overlap integral (\app{a:linover}) should scale as $Q_n\propto\omega_n^{11/6}$, given our normalization convention in \eq{e:norm}.

More recently, \citet{fuller12} showed that excitation can also proceed near the spike in the \brunt{} frequency that occurs at the transition between \co{} and helium in a \cowd{} (see \fig{f:propdiag}). Their corresponding prediction for the torque scaling is $\torq\propto\omegt^5$, implying $Q_n\propto\omega_n^{3}$. Thus this mechanism predicts a steeper overlap scaling with frequency than for excitation at the RCB.

An additional feature of excitation at a composition boundary is that waves originate from a location inside the propagation cavity, meaning both an ingoing and outgoing wave are created. Since the ingoing wave reflects at the inner turning point, interference occurs between the reflected ingoing wave and the purely outgoing wave. Constructive interference implies a large overlap integral $Q_n$, whereas destructive interference makes the overlap small, thus this mechanism predicts a jagged overlap profile with respect to the wave frequency $\omega_n$.

With these theoretical predictions in hand, the essential question to answer is which excitation mechanism---RCB or composition gradient---is most efficient in a given \wdw{} model.\footnote{This question was not addressed in \citet{fuller12} since they adopted an absorbing boundary condition near the outer turning point, and thus did not include the convection zone in their calculations.} The answer hinges on the properties of the convective envelope. \ta{t:wdmods} and \fig{f:propdiag} show that this envelope is very small in hot \wdws{}, and exists at very low densities, but that its extent increases rapidly as a \wdw{} cools. Thus it seems possible that excitation at the RCB may occur for cooler \wdws{}, whereas hotter \wdws{} must rely on the composition gradient mechanism.

\begin{table}
 \caption{\Wdw{} $l=2$ eigenmode properties. Asymptotic fits to numerically computed eigenmode properties for the \wdw{} models from \ta{t:wdmods}. The linear overlap integral $Q_n$ (\app{a:linover} \& \fig{f:Qplots}) is fit as $Q_n = Q_0 (\omega_n/\omegdyn)^a$; the damping rate $\gamma_n$ (\app{a:lindamp} \& \fig{f:gammaplots}) is fit as $\gamma_n = \gamma_0 (\omega_n/\omegdyn)^{-b}$; and the inverse group travel time $\alpha_n = 2\pi/\tgroup{}_{\mr{,} n}$ (\app{a:lindamp}) is fit as $\alpha_n = \alpha_0 (\omega_n/\omegdyn)^c$. $^\dagger$Note that rapid thermal diffusion near the outer turning point causes g-modes of radial order $n\gtrsim50$ to become traveling waves in our \coid{12} and \heid{10} models, meaning our fits for $\gamma_n$ are not relevant in this regime; see \app{a:lindamp} and \fig{f:gammaplots}.}\label{t:eigprops}
\vspace{-1.em}
\begin{center}
\renewcommand{\arraystretch}{1.4}
  \begin{tabular}{l|ll|cc|ll}\hline
 ID & \multicolumn{1}{c}{$Q_0$} & \multicolumn{1}{c|}{$a$} & \multicolumn{1}{c}{$\gamma_0/\omegdyn$} & \multicolumn{1}{c|}{$b$} & \multicolumn{1}{c}{$\alpha_0/\omegdyn$} & \multicolumn{1}{c}{$c$}\\\hline

\heid{10}$^\dagger$ & $9.6\times 10^{-6}$ & 2.61 & $1.5\times 10^{-11}$ & 6.16 & 0.0891 & 2.00\\
\heid{7} & $3.6\times 10^{-6}$ & 1.83 & $2.1\times 10^{-12}$ & 2.00 & 0.158 & 2.00\\
\heid{5} & $7.8\times 10^{-4}$ & 1.90 & $7.7\times 10^{-15}$ & 1.99 & 0.298 & 2.00\\
\hline\coid{12}$^\dagger$ & $7.2\times 10^1$ & 4.40 & $1.2\times 10^{-14}$ & 6.41 & 0.403 & 2.00\\
\coid{6} & $2.7\times 10^1$ & 3.69 & $2.9\times 10^{-14}$ & 1.88 & 0.743 & 2.00
\\\hline
\end{tabular}
\end{center}
\end{table}

One method we can utilize to distinguish between the two mechanisms is simply to observe the power law scaling $Q_n\propto\omega_n^a$ of numerically computed linear overlap integrals for our various \wdw{} models, given in \ta{t:eigprops}. Consistent with our expectations, cooler \hewd{}s have a power law index $a\sim1.83\approx11/6$, implying excitation at the RCB, while hotter \hewd{}s and our \co{} models have larger values of $a$. Furthermore, \fig{f:Qplots} shows that models with steeper overlap power laws also show jagged variation of  $Q_n$ with frequency, thus demonstrating the interference predicted by composition gradient excitation.

Note that if a gravity wave in a cool helium \wdw{} begins to break near its outer turning point (\se{s:wbreak}), this implies that the tidal excitation and wave breaking regions would be almost directly adjacent. It might then be possible for breaking to inhibit excitation, meaning that the composition gradient mechanism would again dominate. A more sophisticated hydrodynamical calculation is required to address this concern.

\subsection{Traveling wave resonance locks}\label{s:trav_intro} \label{s:effres}
The resonance lock scenario we described in \se{s:reslock} relied on resonances between standing \wdw{} eigenmodes and the tidal driving frequency. However, resonance locks are in fact a more general phenomenon that does not explicitly require standing waves.\footnote{\label{fn:resterm}In the traveling wave regime, true ``resonances'' do not occur. Nonetheless, we continue using the term ``resonance lock'' in this context due to the many similarities between standing wave and traveling wave results. In particular, the tidal evolution scenario associated with what we call a traveling wave resonance lock is identical to that associated with a true resonance lock in the standing wave regime, and the transition between standing and traveling wave torques introduced by wave breaking occurs near would-be standing wave resonances.} In the context of \wdw{} binary inspiral, the two essential requirements on the tidal torque function $\torq$ in order for a resonance lock to occur are:
\begin{enumerate}
\item[a)] The torque profile must be a jagged function of the $l=m=2$ tidal driving frequency $\omegt=2(\Omegdiff)$ (\eqp{e:jagged} below).
 \item[b)] The magnitude of the tidal torque must be large enough that it can satisfy \eq{e:reslock2}: $\torq = \Is\Omega/\tgw$.
\end{enumerate}
When these conditions are satisfied and a resonance lock occurs, the tidal quality factor $\Qt$ and heating rate are given by equations \eqref{e:Q} and \eqref{e:Edot}, respectively.

We first address criterion (a). Dropping the tidal energy deposition term from \eq{e:omegdots}, as justified in \se{s:reslock}, yields the simplified orbital evolution equation
\begin{equation}\label{e:simporbevol}
 \frac{1}{m}\frac{d\omegt}{d\Omega} = 1-\frac{\tgw\torq}{\Is\Omega},
\end{equation}
where the gravitational wave decay time $\tgw(\Omega)$ is defined in \eq{e:tgw}.

Let us assume that the tidal torque satisfies criterion (b) at an orbital frequency $\Omega_0$ and a tidal driving frequency $\omegt_0$, so that $d\omegt/d\Omega=0$ and \eq{e:simporbevol} reduces to
\begin{equation}\label{e:reslock3}
 \Is\Omega_0 = \tgw(\Omega_0)\torq(\Omega_0, \omegt_0).
\end{equation}
As long as $\torq$ increases with $\omegt$, \eq{e:reslock3} represents a stable fixed point of the evolution equations; see \fig{f:vecplot}. Next, since the orbital frequency steadily increases due to the emission of gravitational waves, we examine what happens to this fixed point when $\Omega$ changes by a small amount $+\Delta\Omega$. In order to preserve \eq{e:reslock3}, the tidal driving frequency must commensurately change by an amount $\Delta\omegt$ given by
\begin{equation}\label{e:sigchg} 
 \frac{\Delta\omegt}{\omegt} =-\frac{1}{3}\left(\frac{\Delta\Omega}{\Omega}\right)\left( \frac{\partial\log \torq}{\partial\log \omegt} \right)^{-1},
\end{equation}
which can be derived by differentiating \eq{e:reslock3} and substituting \eq{e:tgw}.

\Eq{e:sigchg} allows us to appropriately quantify the ``jagged'' variation of the torque function required by criterion (a): if
\begin{equation}\label{e:jagged} 
 \left|\frac{\partial\log\torq}{\partial\log\omegt}\right|\gg1,
\end{equation} 
then the fixed point can be maintained by only a minimal change in the forcing frequency for a given increase in the orbital frequency, thus constituting a resonance lock. Any general power law trend of $\torq$ with $\omegt$ will fail to satisfy this condition---additional sharp features are required.\footnote{\citet{fuller12} also noticed that $\omegt\approx \mr{constant}$ occurred in their simulations, although they attributed this to the overall power law trend of their torque function with $\omegt$. Indeed, their results possess sharp interference-generated features that provide a much larger contribution to $|d\log \torq/d\log \omegt|$ than the trend, meaning a resonance lock was likely responsible for maintaining $\omegt\approx \mr{constant}$.}

Torque profiles consistent with \eq{e:jagged} can be provided in several ways. For standing waves, the comb of Lorentzians produced by resonances with eigenmodes (see \eqp{e:torque1} and \fig{f:kxplot}) easily satisfies \eq{e:jagged}, since \wdw{} eigenmodes are weakly damped, meaning on- and off-resonance torque values differ by many orders of magnitude. For traveling waves, if the composition gradient mechanism of \citet{fuller12} discussed in \se{s:excite} is the dominant source of wave excitation, it naturally provides sharp features in the torque function due to wave interference. This can also be observed in \fig{f:torqueplot}, where the traveling wave torque changes by a factor of $\sim5$ as $\omegt=2\dOmeg$ changes by only $\sim 10$\%, implying $|d\log \torq/d\log \omegt|\sim 50$.

Lastly, wave breaking can also provide rapid variation in the torque profile due to a sudden transition between standing and traveling wave torques that occurs near resonances at short orbital periods. Specifically, as the tidal driving frequency $\omegt$ sweeps towards a resonance  due to orbital decay by gravitational waves, a tidally excited g-mode's amplitude can become large enough to induce wave breaking (\se{s:wbreak}), which causes the effective damping rate and hence the resulting torque to increase enormously (see the blue curve in \fig{f:torqueplot}).

The precise shape of this transition requires hydrodynamical simulations to ascertain. Fortunately, we find that essentially any transition between a nonresonant standing wave torque in between resonances and a traveling wave torque near resonance will satisfy \eq{e:jagged} for \wdw{}s, due to the large disparity between typical damping times associated with standing waves  and the group travel time, which approximates the damping time for a traveling wave (\ta{t:eigprops} \& \fig{f:gammaplots}). We discuss this further in \se{s:travstand}.

Next, we address criterion (b) for a resonance lock stated at the beginning of this section by estimating the magnitude of the traveling wave torque $\torqtrav$. \citet{goodman98} computed $\torqtrav$ caused by dynamical tides raised in solar-type stars by semi-analytically solving for the traveling wave tidal response.\footnote{\citet{goodman98} explicitly computed the tidal energy deposition rate $\Edtide$; this can be converted to a torque using \eq{e:etau}.} Then, to approximate the effect of discrete resonances, they attached Lorentzian profiles to their formula for $\torqtrav$. We reverse this procedure, and instead approximate $\torqtrav$ by our standing wave formula in the limit that the mode damping time approaches the group travel time. We establish the fidelity of this approximation in \app{a:travver}.

We thus compute $\torqtrav$ by first using \eq{e:torque2} with the tidal driving frequency $\omegt$ set to a particular eigenfrequency $\omega_n$, $\gamma_n$ replaced by $\alpha_n =  2\pi/\tgroup{}_{,n}$ (where $\tgroup$ is the group travel time; see \app{a:lindamp}), and $\detun_n$ set to zero. This yields
\begin{equation}\label{e:travtorque}
\torqtrav(\sigma=\omega_n,\Omega) \sim 4\Es \epsilon^2 Q_n^2\omega_n/\alpha_n,
\end{equation}
where we have approximated $W^2\approx1$. Then, in order to evaluate an effective traveling wave torque for arbitrary $\omegt$, we simply interpolate over values computed using \eq{e:travtorque}.

To estimate the first orbital period $\Ptrl$ at which traveling wave resonance locks can occur, we follow the same procedure as in \se{s:reslock} and again invoke approximate scalings for the eigenmode linear tidal overlap integral $Q$ (\app{a:linover}) and the effective traveling wave damping rate $\alpha$ (\app{a:lindamp}):
\begin{displaymath}
 Q\approx Q_0 (\omegt/\omegdyn)^a\quad\text{and}\quad\alpha\approx\alpha_0(\omegt/\omegdyn)^{c},
\end{displaymath}
where $c=2$; see \ta{t:eigprops} and \fig{f:gammaplots}. The resulting formula, scaled to values for our \coid{6} model (\ta{t:wdmods}), is
\begin{equation}\label{e:Ptrl}
\Ptrl \sim 43\ \mr{ min}\left( \frac{\tdyn}{2.9\ \mr{s}} \right)F_\mr{trl}^q,
\end{equation}
where $\tdyn=(R^3/GM)^{1/2}$ is the \wdw{}'s dynamical time, the factor $F_\mr{trl}$ is
\begin{equation}\begin{split}\label{e:Ftrl}
F_\mr{trl}&\sim\left( \frac{M'}{M} \right)\left( \frac{1+M'/M}{2} \right)^{-5/3}\\
&\times\left( \frac{\betas}{0.010} \right)^{-5}\left( \frac{\Is}{0.18 MR^2} \right)^{-1} \\
&\times\left( \frac{Q_0}{27} \right)^{2}\left( \frac{\alpha_0}{0.74\omegdyn} \right)^{-1}
\\&\times \left(8.41\times10^{12}\right)(0.0119)^{1/q},
\end{split}\end{equation}
the power $q$ is in general
\begin{equation}
 q=\frac{1}{-1/3+2a}<1,
\end{equation}
and $q=0.15$ for our \coid{6} model (\ta{t:eigprops}). (The last line of \eqp{e:Ftrl} is equal to unity for $q=0.15$.) \Eq{e:Ptrl} assumes the \wdw{} begins completely unsynchronized; it is equivalent to equation\ (79) of \citet{fuller12}.

Direct numerical evaluation of eigenmode properties with our \coid{6} \wdw{} model yields $\Ptrl = 40$ min for an equal-mass companion, due to an $n=27$ g-mode, which agrees well with \eq{e:Ptrl}. Values of $\Ptrl$ for each of our fiducial models are provided in \ta{t:tideprops}. Note, however, that in deriving these results for $\Ptrl$ we have assumed that the \wdw{} spin is much smaller than the orbital frequency; if significant synchronization has already occurred, the true value of $\Ptrl$ will deviate from our prediction by an order-unity factor.

\section{Numerical simulations}\label{s:travstand} \label{s:resnum}
\begin{figure}
  \centering
  \includegraphics{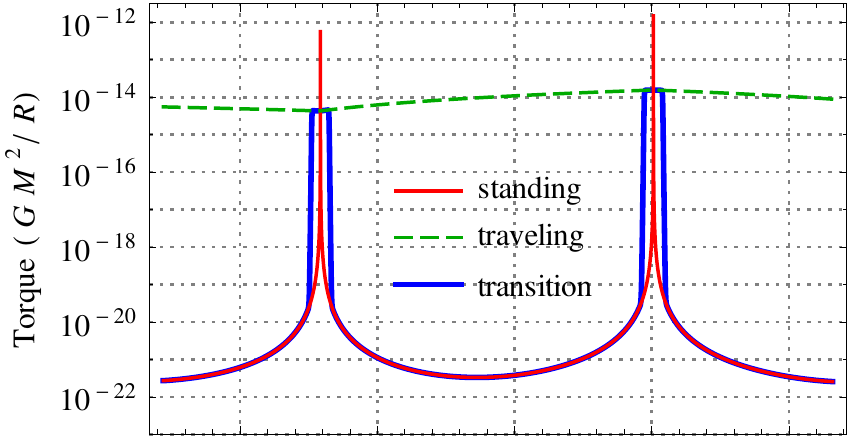}\\
  \includegraphics{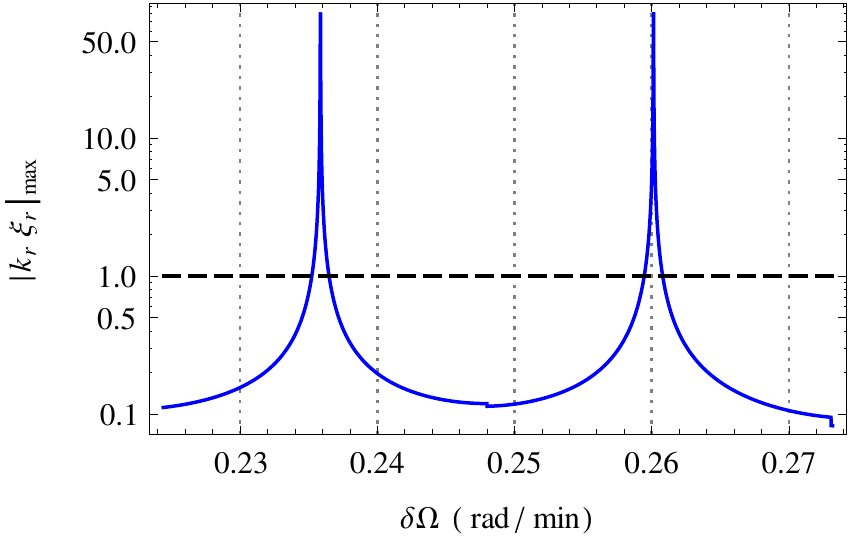}
  \caption{\textit{Top panel:} Example plot of the standing wave torque $\torqstand$ (\eqp{e:torque1}; red line), the traveling wave torque $\torqtrav$ (interpolation over \eqp{e:travtorque} evaluated at eigenmode frequencies; dashed green line), and our interpolation between the two regimes (\eqp{e:torqueinterp}; thick blue line), as functions of $\dOmeg=\Omegdiff$ at fixed $\Porb=30$ min for our \coid{6} model (\ta{t:wdmods}) and an equal-mass companion. The standing wave torque on average is many orders of magnitude smaller than the traveling wave torque; however, near resonances it becomes many orders of magnitude larger. Wave breaking acts to ``cap'' the Lorentzian peaks of the standing wave torque in the interpolation function. \textit{Bottom panel:} Plot of the wave breaking criterion $\krxirabs$ maximized over all eigenmodes and the entire propagation cavity (blue line), using the same parameters and model as the top panel. When $\krxirm<1$, the dynamical tide represents a traveling wave, and the torque $\torqtrans\rightarrow\torqstand$; when $\krxirm>1$, wave breaking occurs, and $\torqtrans\rightarrow\torqtrav$ (\se{s:wbreak}).}
  \label{f:torqueplot}
\end{figure}

\begin{figure*}
  \centering
  \renewcommand{\arraystretch}{0.}
  \begin{tabular}{r|l}
  \multicolumn{1}{c|}{\normalsize\heid{7} \wdw{} model: $\Teff=7,000$ K, $0.2\Msun$} & \multicolumn{1}{c}{\normalsize\coid{6} \wdw{} model: $\Teff=5,500$ K, $0.6\Msun$\rule[-4pt]{0pt}{0pt}}\\\hline
  \includegraphics{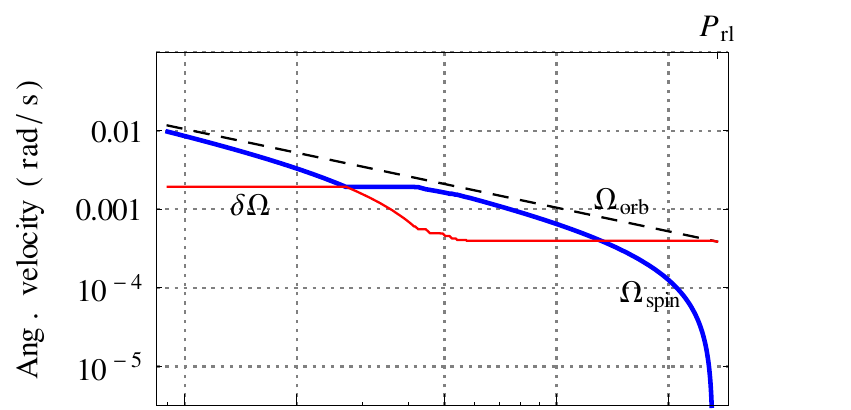} & \includegraphics{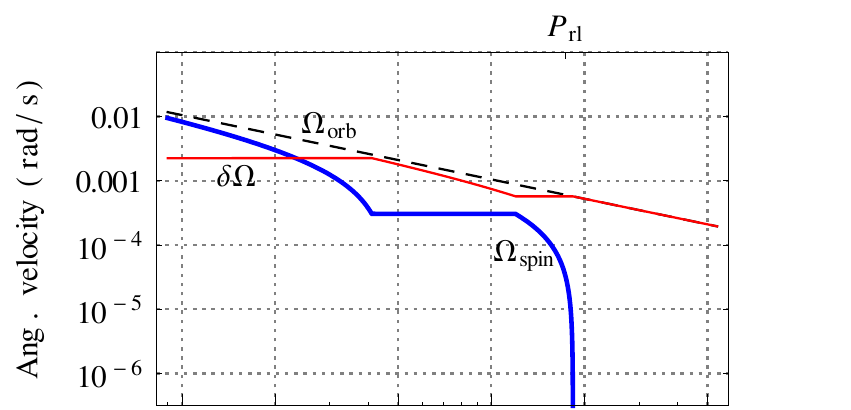} \\
  \includegraphics{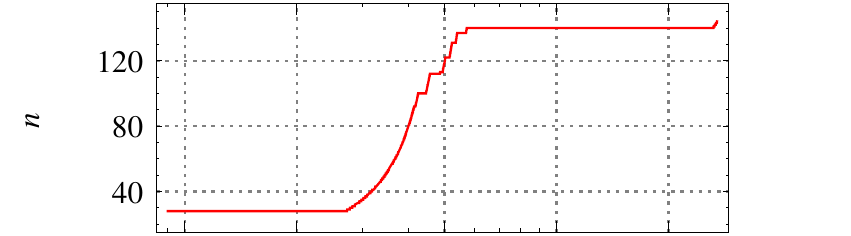} & \includegraphics{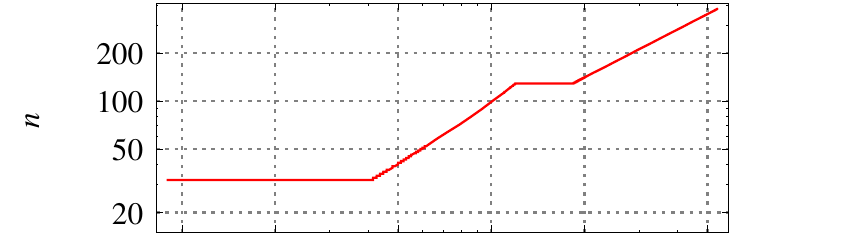} \\
  \includegraphics{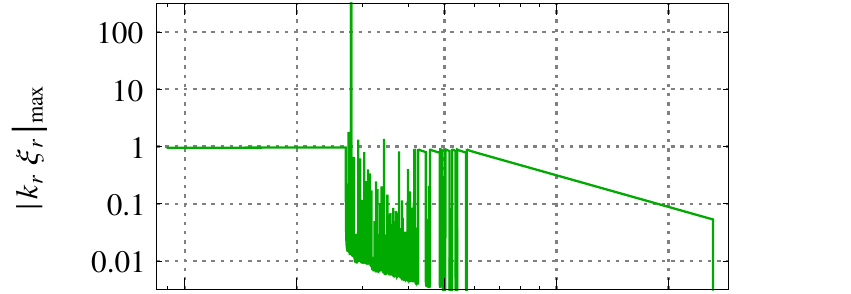} & \includegraphics{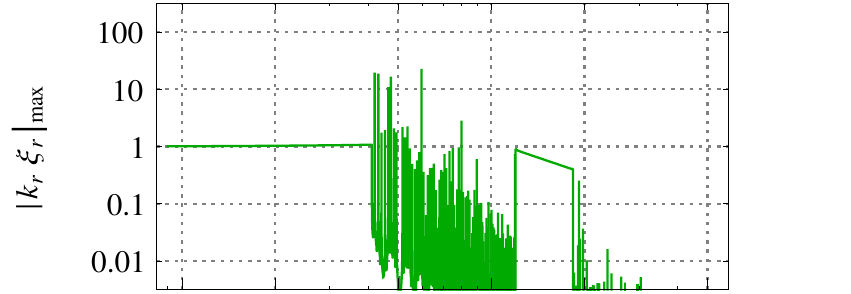} \\
  \includegraphics{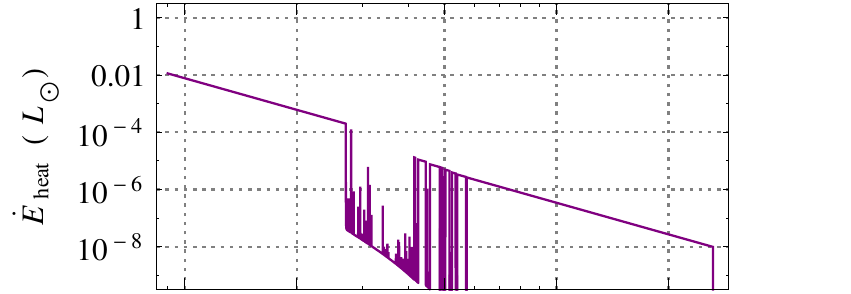} & \includegraphics{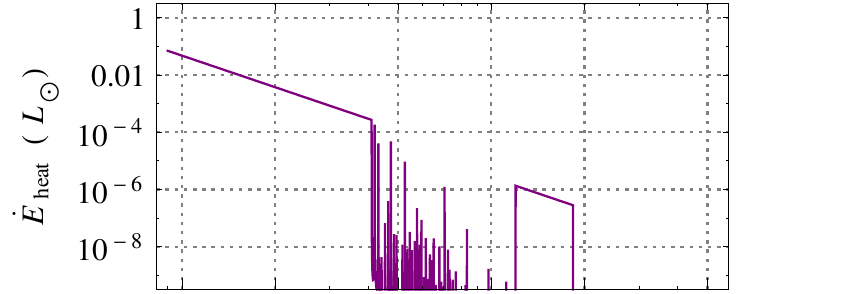} \\
  \includegraphics{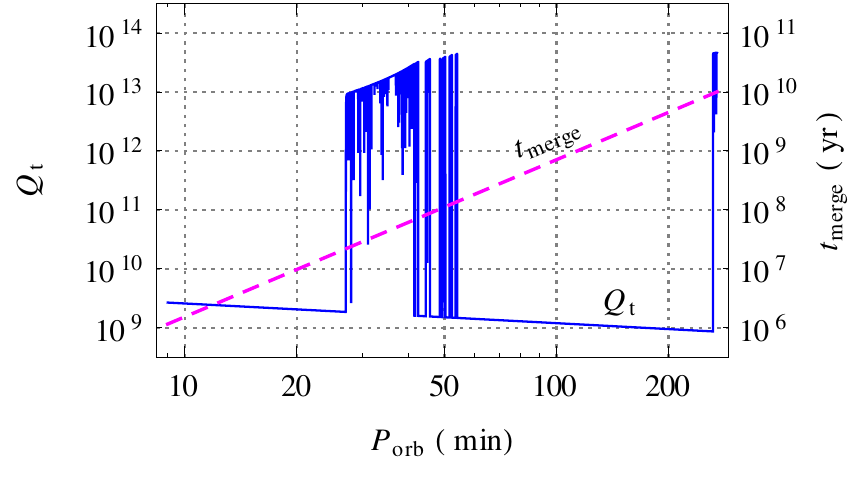} & \includegraphics{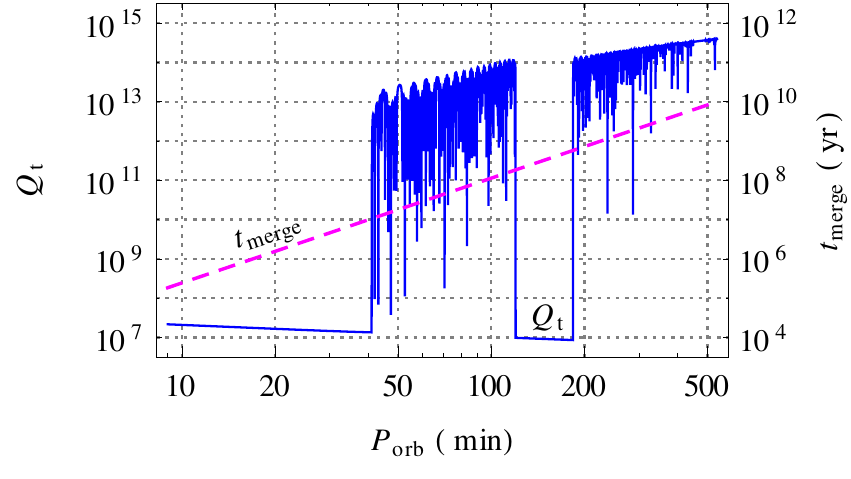} \\
  \end{tabular}
  \caption{Results of numerical simulations of the secular evolution of \wdw{} binaries, the details of which are described in \se{s:resnum}. The left column shows results using our $0.2\Msun$, $\Teff=7,000$ K \heid{7} model, while the right column used our $0.6\Msun$, $\Teff=5,500$ K \coid{6} model (\ta{t:wdmods}). {\it Top row:} The orbital (dashed black line) and spin (thick blue line) frequencies, as well as their difference $\dOmeg=\Omega-\Omegsp$ (red line); the latter sets the tidal forcing frequency $\omegt=2\dOmeg$. Resonance locks correspond to regions where $\dOmeg$ is constant. Our assumption of slow rotation breaks down when $\dOmeg\lesssim\Omegsp$ due to nonlinear rotational modification of stellar eigenmodes. {\it Second row:} Number of radial nodes $n$ of dominant eigenmode/wave. {\it Third row:} Maximum value of $\krxirabs$ across entire \wdw{}, evaluated assuming standing waves, which assesses whether wave breaking occurs (\se{s:wbreak}). During the initial resonance lock, $\krxirm$ starts $<1$, but gradually rises until it becomes $\sim 1$ and breaking begins. {\it Fourth row:} Rate at which orbital energy is dissipated as heat in the \wdw{}, in units of $L_\sun$. {\it Bottom row:} Tidal quality factor $\Qt$ (blue line) and time until mass transfer $\tinsp=3\tgw/8$ (\eqp{e:tgw}; dashed magenta line). See equations \eqref{e:Q} and \eqref{e:Edot} for analytic estimates of the tidal quality factor $\Qt$ and heating rate, respectively.}
  \label{f:rlockco}
\end{figure*}

To address the tidal evolution of an inspiraling \wdw{} binary undergoing resonance locks, we aim to combine the standing and traveling wave results from \S\S{} \ref{s:reslock} \& \ref{s:trav} numerically. To this end, we evaluate the complete standing wave tidal torque from \eq{e:torque1} and solve for the spin and orbital evolution using \eq{e:omegdots}. To account for wave breaking, we check that all eigenmodes satisfy $\krxirm<1$ throughout the \wdw{} (\se{s:wbreak}); when an eigenmode exceeds unit shear, we instead set its damping rate to $\alpha_n=2\pi/\tgroup{}_{,n}$ (\app{a:lindamp}), which approximates the traveling wave regime \citep{goodman98}. We smoothly transition between the standing and traveling wave regimes using the interpolation formula
\begin{equation}\label{e:torqueinterp} 
 \torqtrans = \frac{\torqstand +  \torqtrav\left(\krxirm\right)^z}{1+\left(\krxirm\right)^z},
\end{equation} 
where $\torqstand$ is the standing wave torque from \eq{e:torque1}, $\torqtrav$ is the traveling wave torque produced by interpolating over \eq{e:travtorque}, and $\krxirm$ is the maximum value of the wave shear over all relevant eigenmodes and across the entire propagation cavity (\se{s:wbreak}), evaluated assuming standing waves. We arbitrarily adopt $z=25$ to induce a sharp transition that occurs only when $\krxirm$ is very close to 1; our results are insensitive to the value of $z$ so long as it is $\gtrsim |\ln(\torqtrav/\torqstand)|$. \fig{f:torqueplot} shows a comparison of $\torqstand$, $\torqtrav$, and the transition function in \eq{e:torqueinterp}.

\fig{f:rlockco} shows the results of two of our simulations. The left column used our  $0.2\Msun$, $\Teff=7,000$~K \heid{7} model, while the right column used our $0.6\Msun$, $\Teff=5,500$~K \coid{6} model (\ta{t:wdmods}). We did not account for \wdw{} cooling or tidal heating, and instead used fixed \wdw{} models throughout both simulations. We initialized our simulations with $\Omegsp\sim0$, and the orbital period set so that the time until mass transfer $\tinsp=3\tgw/8$ (\eqp{e:tgw}) was equal to 10 billion years.

Both simulations follow the archetypal scenario laid out in \se{s:dynregimes}, transitioning amongst the four regimes \SID1, \SID2, \TID1, and \TID2. Both begin in \SID1, where the dynamical tide is a standing wave even near resonances, but the tidal torque is too weak to create a resonance lock. As the orbit shrinks due to gravitational wave radiation, the tidal force waxes and the first resonance lock eventually begins in both simulations at the appropriate value of $\Prl$ estimated in \se{s:reslock} and provided in \ta{t:tideprops}; this is regime \SID2. At this point tidal heating and synchronization suddenly become much more efficient (\se{s:energ}), and the difference between orbital and spin frequencies remains constant. Analytic formulas for the tidal quality factor and heating rate appropriate for this situation (as well as \TIDnp2 discussed below) are given in equations \eqref{e:Q} and \eqref{e:Edot}, respectively, and exactly reproduce their numerically derived values appearing in the bottom two rows of \fig{f:rlockco}.

Both simulations begin the standing wave resonance lock regime \SID2 with a value of the wave breaking criterion $\krxirm<1$; however, as the orbit shrinks further, progressively larger and larger wave amplitudes become necessary to support a resonance lock, eventually leading to wave breaking near the outer turning point (\se{s:wbreak}). At the onset of \SID2 in the \coid6 simulation, the inequality $\krxirm<1$ is only weakly satisfied, meaning that the standing wave lock regime \SID2 is short lived. In the \heid7 simulation, however, \SID2 begins with $\krxirm\ll1$, so that the initial resonance lock persists from $\Prl=270$ min to a period of $P\approx 50$ min, corresponding to an interval of time of about 5 billion years.

Once $\krxirm$ becomes $\sim1$, both simulations enter regime \TID1, where near would-be resonances the dynamical tide becomes a traveling wave too weak to create a resonance lock. The otherwise steeply peaked standing wave torque is thus capped in this regime; see \fig{f:torqueplot}. Regime \TID1 results in a weak tidal synchronization and heating scenario, very similar to \SID1.

Eventually, at an orbital period $\sim\Ptrl$ (\se{s:effres}; \ta{t:tideprops}), both simulations enter regime \TID2, where even the traveling wave torque can create a resonance lock (terminology discussed further in footnote \ref{fn:resterm}). Tides again become efficient, with synchronization and heating scenarios quantitatively consistent with the analytic results in \se{s:energ} (just as in \SIDnp2). In the \heid7 simulation, \TID2 begins at an orbital period of $\approx 27$ min, which differs from its value of $\Ptrl=49$ min listed in \ta{t:tideprops}, since that value is only strictly applicable when $\Omegsp=0$, whereas significant synchronization has already occurred. The value of $\Ptrl$ in the \coid6 simulation is a better estimate of the onset of \TID2 due to the brief duration of \SID2 in that case.

The maximum wave shear $\krxirm$ shown in \fig{f:rlockco} (which is evaluated assuming standing waves) remains very close to unity throughout much of regime \TID2. A reasonable question, then, is whether this is an artifact of the interpolation function we used to transition between standing and traveling waves torques (\eqp{e:torqueinterp}).

On the contrary, we believe there is a physical reason why $\krxirm$ should saturate at $\sim1$, and that it is a natural consequence of the traveling wave resonance lock scenario we proposed in \se{s:trav_intro}. Specifically, at this point in the system's evolution, if the dynamical tide attempts to set up a standing wave, the orbital frequency will evolve, increasing the tidal driving frequency $\omegt=2(\Omegdiff)$ towards a resonance and inducing wave breaking. However, fully transitioning to the traveling wave regime then creates a much larger torque (due to the much larger effective damping rate), causing the spin frequency to increase rapidly and sending $\omegt$ away from resonance, ending wave breaking and reinstituting the standing wave regime. The end result is that $\krxirm$ should average to be $\sim1$.

This line of reasoning suggests that the true phenomenon may be episodic in nature. Alternatively, a weak-breaking regime may be possible, allowing the system to smoothly skirt the boundary between linear and nonlinear fluid dynamics. Full hydrodynamical simulations may be necessary to understand this in more detail.

\section{Discussion}\label{s:discuss} 
\subsection{Observational constraints}\label{s:discobs} 
\newcommand{\Jsys}{J0651}
The theoretical results we have developed can be compared to the recently discovered system SDSS J065133.33+284423.3 (henceforth \Jsys), which consists of a $\Teff=16,500$ K, $0.26\Msun$ \hewd{} in a 13-minute eclipsing binary with a $\Teff=8,700$ K, $0.50\Msun$ \cowd{} \citep{brown11}. Orbital decay in this system consistent with the general relativistic prediction was discovered by \citet{hermes12}.

\citet{piro11} studied tidal interactions in \Jsys, and produced lower limits on values of the tidal quality factor $\Qt$ by assuming that the observed luminosity of each \wdw{} is generated entirely by tidal heating, and that both \wdw{}s are nonrotating. However, the definition of $\Qt$ used in that work differs with ours (\eqp{e:Qdef}), which hinders  straightforward comparison.\footnote{The relationship between our value of the tidal quality factor, $\Qt$, and that used in \citet{piro11}, $\Qt'$, is $\Qt'=\Qt\,\omegt M/\lambda\Omega\mu$, where $\mu$ is the reduced mass. Our value $\Qt$ is consistent with being the reciprocal of an effective tidal lag angle, which is the conventional definition; see \se{s:Q}.}

Instead, we compare the observed luminosities with our expression for $\Edheat$ from \eq{e:Edot}, which is applicable during a resonance lock. All parameters for \Jsys{} entering into \eqref{e:Edot} were determined observationally except the moments of inertia $\Is$ and the asynchronicity periods $\Pasync=2\pi/\dOmeg$, where $\dOmeg=\Omegdiff$. Note that \eq{e:Edot} counterintuitively shows that greater synchronization leads to diminished tidal heating, since the heating rate is proportional to the degree of synchronization (and hence inversely proportional to $\Pasync$). We can thus use appropriate values of $\Is$ from \ta{t:wdmods} and impose the inequality $L\gtrsim\Edheat$, since cooling can also contribute to each luminosity, in order to constrain $\Pasync$ for each \wdw{}.

This calculation yields $\Pasync\gtrsim7$ min for the \hewd{} and $\Pasync\gtrsim400$ min for the \cowd{}, each with uncertainties of $\sim20$\%. Since the orbital period of \Jsys{} is such that resonance locks should currently exist in both \wdw{}s---$\Porb$ is less than both $\Prl$ from \se{s:reslock} for standing waves and $\Ptrl$ from \se{s:trav_intro} for traveling waves (\ta{t:tideprops})---and since our simulations developed wave breaking long before $\Porb=13$~min (\fig{f:rlockco}), our \emph{a priori} expectation is that $\Pasync$ should be $\sim\Ptrl\sim50$~min for each \wdw{}. It is encouraging that the inferred constraints on $\Pasync$ for both \wdws{} are within an order of magnitude of this prediction.

We can nonetheless comment on what the deviations from our predictions may imply. First, the fact that $\Pasync > 7\ \mr{min}<\Porb=13$ min for the \hewd{} means that explaining its luminosity purely by tidal heating would require retrograde rotation. Since this situation would be highly inconsistent with our results, we can conclude that its luminosity must be generated primarily by standard \wdw{} cooling or residual nuclear burning rather than tidal heating. If this is correct, it would imply an age for the \hewd{} of only $\sim 40$ Myr \citep{panei07}; dividing this age by a cooling time of $\sim1$~Gyr (\ta{t:wdmods}) yields a very rough probability for finding such a system of $\sim4$\%. This scenario does not seem unlikely, however, since selection bias favors younger \wdw{}s.

On the other hand, the inferred lower limit of $\sim400$ min placed on $\Pasync$ for the \cowd{} is much larger than our predictions for both $\Prl$ and $\Ptrl$.  Furthermore, we find that tidal heat is deposited very close to the photosphere in hot \cowd{} models (\ta{t:wdmods}), so we expect tidal heating to contribute directly to the luminosity of the \cowd{} in \Jsys{} (\se{s:heat}). The constraint on $\Pasync$ thus means that the \cowd{} appears to be more synchronized than our theoretical expectation, and consequently less luminous than our prediction by a factor of $\sim 400\; \mr{min}/\Ptrl\sim10$.

Although this is formally inconsistent with our results, examining the first row of \fig{f:rlockco} shows that both of our numerical simulations have $\dOmeg\ll\Omegsp$ near $\Porb=13$ min. This means that the influence of rotation on eigenmode properties is likely to be very important at such short orbital periods (\se{s:nonlinrot}), which is not included in our analysis. This could lead to enhanced synchronization and hence mollify the discrepancy (since, again, increased synchronization implies less tidal heating). Damping and excitation of \wdw{} eigenmodes by nonlinear processes are also likely to be important considerations, which could also increase the efficiency of tidal synchronization.

Lastly, assuming resonance locks are occurring in both \wdw{}s, we predict that  the rate of orbital decay should be enhanced due to tides by (\se{s:Pdot})
\begin{displaymath}
 \left(\frac{\Pdottide}{\Pdotgw}\right)_\mr{\Jsys}\!\!\!\sim3\%,
\end{displaymath}
where $\Pdotgw=-\Porb/\tgw$. Although this estimate fails to include the effect of rotation on eigenmode frequencies, which we already argued may be important in \Jsys, it should nonetheless be robust at the order-of-magnitude level. This $\sim3\%$ deviation between the system's period derivative and the general relativistic prediction not accounting for tides may be detectable given further sustained observations \citep{piro11}.

\subsection{Rotation and \wdw{} evolution}\label{s:nonlinrot}
In our analysis we have neglected the influence of rotation on the stellar eigenmodes beyond the simple geometrical Doppler shift of the forcing frequency into the corotating frame. To linear order in the rotation frequency, the correction to the stellar eigenfrequencies makes very little difference to the results we have derived---it just means there should be factors of $(1-C_n) \sim5/6$ appearing in various formulas in \se{s:reslock}, which we neglected for simplicity.

However, when $\dOmeg=\Omegdiff\lesssim\Omegsp$, nonlinear rotational effects become important. \fig{f:rlockco} shows that this inequality is satisfied below $\Porb\sim25$ min in our \coid6 simulation,  and takes hold soon after the first resonance lock in our \heid7 simulation, at only $\Porb\sim150$ min. Below these orbital periods, fully accounting for the Coriolis force in the stellar oscillation equations becomes necessary.

For example, excitation of rotationally supported modes---Rossby waves and inertial waves---could prove very efficient. Such modes have corotating-frame eigenfrequencies that are strongly dependent on the rotation frequency, so a resonance lock would follow the more general trajectory \citep{witte99}
\begin{equation}\label{e:reslock4} 
 0=\dot{\detun}=m\left[ \left( 1 + \frac{1}{m}\frac{\partial\omega_n}{\partial\Omegsp} \right)\Omegspd - \Omegd \right].
\end{equation}
Since our analysis in \se{s:energ} relied on resonance locks producing $\Omegd\approx\Omegspd$, which no longer holds when $\partial\omega_n/\partial\Omegsp\ne0$, it is unclear whether nonlinear rotational effects could substantially alter our results for e.g.\ the tidal quality factor (\eqp{e:Q}) and tidal heating rate (\eqp{e:Edot}).

\label{s:discheat} 
\Wdw{} cooling and tidal heating could also potentially modify the synchronization trajectory that results during a resonance lock. For example, since g-mode frequencies approximately satisfy $\omega_{nl}\sim\langle N\rangle l/n$, and since the \brunt{} frequency scales with temperature as $N\propto T^{1/2}$ in a degenerate environment (\se{s:prelim}), progressive changes in a \wdw{}'s thermal structure due to either heating or cooling would introduce an additional $\partial\omega_n/\partial t$ term on the right-hand side of \eq{e:reslock4}.

\subsection{Crystallization}\label{s:crystal}
Whether a plasma begins to crystallize due to ion-ion electromagnetic interactions is determined by the Coulomb interaction parameter $\Gamma$, which is defined as the ratio of the Coulomb to thermal energy,
\begin{equation}
 \Gamma = \frac{Z^2e^2}{d_\mr{i} kT},
\end{equation}
where $Ze$ is the mean ion charge and $d_\mr{i}$ is the ion separation, defined by $1=n_\mr{i}(4\pi/3)d_\mr{i}^3$. When $\Gamma \gtrsim1$, the plasma under consideration behaves as a liquid; when $\Gamma > \Gamcrys$, the plasma crystallizes. This critical value is $\Gamcrys\sim 175$ in single-component plasmas \citep{dewitt01}. However, more recent observational studies of \cowd{} populations  \citep{winget09} as well as detailed theoretical simulations \citep{horowitz07} indicate that a larger value of $\Gamcrys\sim220$ is applicable for two-component plasmas, as in the cores of \cowd{}s.

As shown in \ta{t:wdmods}, the central value of $\Gamma$  does not exceed the appropriate value of $\Gamcrys$ for any of our \hewd{}s. However, for \coid{6}, our $0.6\Msun$, $\Teff=5,500$~K \cowd{}, we have $\Gamma_\mr{core}=260>\Gamcrys$, and further $\Gamma>\Gamcrys$ for 19\% of the model by mass (taking $\Gamcrys=220$). This is indicated in the bottom panel of \fig{f:propdiag} as a shaded region.

The excitation of dynamical tides in \wdws{} possessing crystalline cores is an interesting problem that deserves further study. We will only speculate here on the possible physical picture. Our preliminary calculations of wave propagation inside the crystalline core, using expressions for the shear modulus of a Coulomb crystal from \citet{hansen79}, indicate that the shear wave Lamb frequency is several orders of magnitude too large to allow gravity waves to propagate as shear waves in the core.

Thus it seems possible that dynamical tides could be efficiently excited at the edge of the core, as in excitation at the edge of a convective core in early-type stars \citep{zahn75}. In this scenario the deviation of the tidal response inside the crystal from the potential-filling equilibrium tide solution excites outward-propagating g-modes. As a consequence, tidal gravity waves may be much more efficiently excited in crystalline core \cowd{}s, since the \brunt{} frequency gradient near the core would be much steeper than in any composition gradient zone, as discussed in \se{s:excite}.

\subsection{Nonlinear damping}\label{s:nl} 
Since eigenmodes responsible for resonance locks in the standing wave regime attain large amplitudes, it is natural to worry that they might be unstable to global nonlinear damping by the parametric instability, even if they don't experience wave breaking (e.g.\ \citealt{arras03,weinberg12}). To address this, we performed a rough estimate of the threshold amplitude $T$ for the parametric instability to begin sapping energy from the eigenmode responsible for the standing wave resonance lock in our \coid6 simulation from \se{s:travstand}. Using the procedure detailed in \S~6.5 of \citet{burkart12}, we found $T\sim2\times10^{-8}$. On the other hand, during the resonance lock the mode in question began with an amplitude of  $|q|\sim4\times10^{-8}$, which grew to $\sim10^{-7}$ before wave breaking destroyed the lock.

This demonstrates that parametric instabilities may limit the achievable amplitudes of standing waves in close \wdw{} binaries, potentially somewhat more stringently than wave breaking alone. Exactly how this affects the overall tidal synchronization scenario requires more detailed study.

\section{Conclusion}\label{s:conc}
In this paper, we have studied the linear excitation of dynamical tides in \wdw{} binaries inspiraling subject to gravitational wave radiation. We showed that the phenomenon of resonance locks occurs generically in this scenario, both when the dynamical tide represents a standing wave or a traveling wave. (Our choice of terminology is discussed further in footnote \ref{fn:resterm}.)

In a resonance lock, as the orbital frequency increases according to $\Omegd=\Omega/\tgw$, where $\tgw$ is the gravitational wave inspiral time (\eqp{e:tgw}), a synchronizing torque produced by the dynamical tide causes the \wdw{} spin frequency to evolve at nearly the same rate: $\Omegspd\approx\Omegd$ (\se{s:reslock}). This means the $l=m=2$ tidal driving frequency $\omegt=2(\Omegdiff)$ remains constant, which in turn keeps the tidal torque nearly constant, leading to a stable situation. In other words, a resonance lock is a dynamical attractor (\fig{f:vecplot}).

We first considered resonance locks created by standing waves, where resonances between the tidal driving frequency and \wdw{} eigenmodes create the synchronizing torque required to maintain $\sigma\approx$ constant. We derived analytic estimates of the orbital period $\Prl$ at which such resonance locks can first occur (\se{s:reslock}; also \ta{t:tideprops}): $\Prl\sim30$ min for hot \cowd{}s ($\Teff\sim12,000$~K) and $\Prl\sim200$ min for cold \co{} \wdw{}s ($\Teff\sim6,000$~K). For helium \wdws, we found $\Prl\sim70$ min for hot models ($\Teff\sim10,000$~K), and $\Prl\sim1$~day for colder models ($\Teff\sim5,000$~K).

Tides preferentially deposit orbital angular momentum into a \wdw's outermost layers, where wave damping is most efficient. A concern thus exists that a synchronously rotating critical layer might develop, causing rapid wave damping and eliminating the possibility of maintaining a standing wave \citep{goldreich89b}. However, we showed that critical layers are in fact unlikely to develop in the standing wave regime of \wdw{} binary inspiral, since a typical \wdw{} fossil magnetic field is capable of winding up and enforcing solid-body rotation throughout the \wdw{} down to orbital periods of $\sim10$ min or less (\se{s:reslockval}; \app{a:trans}).

We derived analytic formulas for the tidal quality factor $\Qt$ (\eqp{e:Q}) and heating rate $\Edheat$ (\eqp{e:Edot}) during a resonance lock (\se{s:energ}). (Since $\Qt$ parametrizes the total tidal energy transfer rate, including mechanical energy associated with changing the \wdw{} spin, values of $\Qt$ alone do not determine the tidal heating rate.) Our results predict that, for orbital periods of $\lesssim$ hours, $\Qt\sim10^7$ for \cowd{}s and $\Qt\sim10^9$ for \hewd{}s. Our formula for $\Qt$ is independent of \wdw{} eigenmode properties and weakly dependent on the orbital period, scaling as $\Qt\propto\Porb^{-1/3}$. It is, however, strongly dependent on the \wdw{} mass and radius. We also found that tidal heating begins to rival typical \wdw{} luminosities for $\Porb\lesssim10$ min, a result that is relatively insensitive to \wdw{} properties due to the steep power law scaling $\Edheat\propsim\Porb^{-11/3}$. The analytic results we derived can easily be incorporated into population synthesis models for the evolution of close \wdw{} binaries.

As a standing wave resonance lock proceeds, the wave amplitude required to maintain synchronization grows. Eventually, the amplitude becomes so large that the standing wave begins to break near the surface convection zone (\se{s:wbreak}). This causes the dynamical tide to become a traveling wave, eliminating the resonance lock. This occurred soon after the initial resonance lock in our $0.6M_\sun$, $\Teff=5,500$ K \cowd{} simulation; however, the standing wave resonance lock lasted much longer in our $0.2M_\sun$, $\Teff=7,000$ K \hewd{} simulation, from $\Porb\sim250$ min down to $\sim40$~min, amounting to $\sim10$~Gyr of binary evolution.

Resonance locks have traditionally been considered only when the dynamical tide represents a standing wave \citep{witte99}. We showed, however, that given sufficiently short orbital periods, resonance locks can even occur in the traveling wave regime (\se{s:trav_intro}). We derived two simple criteria for whether traveling waves can effect resonance locks: the traveling wave torque must be large enough to enforce $\Omegd\approx\Omegspd$, and the torque profile as a function of the tidal driving frequency $\omegt=2(\Omegdiff)$ must possess ``jagged'' features, a concept quantified by $|d\log\torq/d\log\omegt|\gg1$ (\eqp{e:jagged}), where $\torq$ is the tidal torque.

The first criterion is satisfied for orbital periods below a critical period $\Ptrl$, which we found to be $\Ptrl\sim40$ -- 50 min in most \wdw{} models (\eqp{e:Ptrl}; \ta{t:tideprops}). The second criterion can be satisfied by rapid transitions between standing and traveling wave torques (which differ by orders of magnitude) near resonances as a result of wave breaking (\se{s:wbreak}), or by wave interference due to excitation by a composition gradient (\se{s:excite}; \citealt{fuller12}). Excitation likely proceeds at a composition gradient in \cowd{}s and hot \hewd{}s, but excitation at the radiative-convective boundary becomes important for colder \hewd{}s with larger surface convection zones (\se{s:excite}). Excitation off a crystalline core may also be important in cold \cowd{}s (\se{s:crystal}).

Even after the initial standing wave resonance lock is destroyed by wave breaking, a new traveling wave resonance lock takes hold once the orbital period declines to $\Porb\sim\Ptrl\sim 40$ -- 50 min. The synchronization trajectory and corresponding values of the tidal quality factor (\eqp{e:Q}) and tidal heating rate (\eqp{e:Edot}) are the same during a traveling wave resonance lock. We confirmed our analytic derivations with numerical simulations that smoothly switched between standing and traveling wave torques based on the maximal value of the wave shear $\krxirabs$ (\fig{f:torqueplot}), with wave breaking leading to traveling waves for $\krxirabs\gtrsim1$ (\se{s:wbreak}). We presented the results of two simulations, one with a \hewd{} and one with a \cowd{}, in \se{s:travstand}. Once the traveling wave resonance lock began, synchronization in our numerical calculations proceeded until the spin frequency $\Omegsp$ became larger than $\dOmeg=\Omegdiff$, meaning nonlinear rotational effects not included in our analysis were likely to be important (\se{s:nonlinrot}).

Our numerical calculations (\fig{f:rlockco}) demonstrate that efficient tidal dissipation is produced by standing wave resonance locks at large orbital periods,  and by traveling wave resonance locks at smaller orbital periods.   The importance of the standing wave resonance lock regime at large orbital periods can be tested by measuring the rotation rates of wide \wdw{} binaries.  We predict that systems with orbital periods of hours should have undergone significant synchronization (\fig{f:rlockco}), while models that focus solely on excitation of traveling waves \citep{fuller12} would predict synchronization only at significantly shorter orbital periods.   A second prediction of our model is that there may be a range of intermediate orbital periods (e.g., $20\ \mr{min}\lesssim  \Porb \lesssim 40\ \mr{min}$) where tidal dissipation is relatively inefficient compared to both smaller and somewhat larger orbital periods.

The results derived here can be directly compared to the recently discovered 13-minute \wdw{} binary \Jsys{} (\se{s:discobs}).  We predict a $\sim3\%$ deviation of the orbital decay rate from the purely general relativistic value, which may be measurable given further observations. We also find that our predicted tidal heating rates are within an order of magnitude of the observed luminosities.   This broad agreement is encouraging given the well-known difficulties tidal theory has accurately predicting the efficiency of tidal dissipation in many stellar and planetary systems.

In detail, we find that even if the \hewd{} is nonrotating (which maximizes the tidal energy dissipated as heat), tidal heating is a factor of $\sim2$ less than the observed luminosity, strongly suggesting that much of its luminosity must derive from residual nuclear burning or cooling of thermal energy rather than tidal heating. In contrast, we predict that the \cowd{} in \Jsys{} should be $\sim10$ times more luminous than is observed.   We suspect that the origin of this discrepancy is the importance of rotational modification of stellar eigenmodes at the short orbital period present in \Jsys{} (\se{s:nonlinrot}), and perhaps the effects of nonlinear damping/excitation of stellar oscillations (e.g.\ \se{s:nl}; \citealt{weinberg12}).   These will be studied in future work.

\acknowledgments
J.B. thanks Jieun Choi for assistance in the preparation of this manuscript. This work was supported by NSF AST-0908873 and NASA NNX09AF98G. J.B. is an NSF Graduate Research Fellow. E.Q. was supported by a Simons Investigator award from the Simons Foundation, the David and Lucile Packard Foundation, and the Thomas Alison Schneider Chair in Physics at UC Berkeley. P.A. is an Alfred P. Sloan Fellow, and received support from the Fund for Excellence in Science and Technology from the University of Virginia.
\begin{appendix}
\onecolumn

\section{Angular momentum transport}\label{a:trans}
\newcommand{\Bfoss}{B_0}
\newcommand{\Nw}{N_\mr{w}}

If we assume there is a source of angular momentum near the \wdw{} surface, e.g.\ from tides, a fossil magnetic field of initial magnitude $\sim \Bfoss$ will wind up and exert magnetic tension forces attempting to enforce solid-body rotation. The rate at which magnetic tension transports polar angular momentum through a spherical surface $S$ at radius $r$ is given by
\begin{equation}\label{e:Jdot}
 \dot{J}_z = \frac{1}{4\pi}\int_S B_r B_\phi \,r \sin\theta \,dS,
\end{equation}
which can be derived by applying the divergence theorem to the magnetic tension force density $(\vec{B}\cdot\grad) \vec{B}/4\pi$. As the field winds up, the radial component remains constant, while the azimuthal component increases as \citep{spruit99}
\begin{equation}
 B_\phi = \Nw B_r,
\end{equation}
where $\Nw = r\sin\theta\int(d\Omegsp/dr)dt$ is the rotational displacement that occurs during wind up and we have assumed the rotational velocity field can be described by ``shellular'' rotation: $\vec{v} = \Omegsp(r) r \sin\theta \hat{\phi}$.

Once the field wind up propagates into the \wdw{} core, an equilibrium field is established that communicates the tidal torque throughout the \wdw{} and eliminates any rotational shear. There are several requirements necessary for this equilibrium to be reached during inspiral, and consequently for \wdws{} in inspiraling binaries to rotate as a rigid bodies.

\subsection{Solid-body rotation at short orbital periods}\label{a:sbody} 
\newcommand{\FA}{F_1}
\newcommand{\FB}{F_2}
\newcommand{\FC}{F_3}
\newcommand{\FD}{F_4}

Whether there is sufficient time for the global equilibrium magnetic field to develop and eliminate differential rotation altogether amounts to whether the timescale over which the torque changes, given by the gravitational wave inspiral time $\tgw$ (\eqp{e:tgw}), is longer than the timescale over which the wind up of the magnetic field propagates into the core, which is given by the global Alfv\'en crossing time $\langle t_\mr{A}\rangle = \int dr/v_\mr{A}$, where $v_\mr{A} = B_r / \sqrt{4\pi\rho}$ is the radial Alfv\'en speed. This restriction translates to
\begin{equation}\label{e:alfvenr1}
 \Bfoss \gg \frac{1}{\tgw}\int_0^R \sqrt{4\pi\rho}\;dr,
\end{equation}
where we have assumed $B_r\sim\Bfoss\sim\text{constant}$. We can evaluate this further as
\begin{equation}\label{e:alfvenr2}
 \Bfoss\gg 0.2\ \mr{G}\;\left( \frac{\Porb}{10\ \mr{min}} \right)^{-8/3}\FA
\end{equation}
where
\begin{equation*}
 \FA=\left( \frac{M'}{M} \right)\left( \frac{1+M'/M}{2} \right)^{-1/3}\left( \frac{M}{0.60\Msun} \right)^{13/6}\left( \frac{R}{0.013\Rsun} \right)^{-1/2}.
\end{equation*} 

The other requirement for solid-body rotation is that the rotational displacement $\Nw$ required for the equilibrium field configuration must not be too extreme, since the field becomes susceptible to resistive dissipation as well as various instabilities as it winds up \citep{spruit99}. We can address this constraint by setting the tidal torque appropriate for a resonance lock $\torq_r = I_r\Omega/\tgw$ (\eqp{e:reslock2}), where $I_r$ is the moment of inertia up to a radius $r$, equal to \eq{e:Jdot}, and then requiring $\Nw\lesssim1$. Solving for $\Bfoss$, we have
\begin{equation}\label{e:windupr1} 
 \Bfoss \gtrsim \sqrt{\frac{\torq}{r^3}} = \sqrt{\frac{I_r\Omega}{r^3\tgw}}.
\end{equation}
Since the right-hand side of \eq{e:windupr1} scales radially as $r^1$, we can safely set $I_r=\Is$ and $r=R$. Evaluating this further yields
\begin{equation}\label{e:windupr2}
 \Bfoss\gtrsim 10^4\ \mr{G}\left( \frac{\Porb}{10\ \mr{min}} \right)^{-11/6}\FB,
\end{equation}
where
\begin{equation}
 \FB = \left( \frac{M'}{M} \right)^{1/2}\left( \frac{1+M'/M}{2} \right)^{-1/6}\left( \frac{\Is}{0.18MR^2} \right)^{1/2}\left( \frac{M}{0.6\Msun} \right)^{4/3}\left( \frac{R}{0.013\Rsun} \right)^{-1/2}.
\end{equation}

We see that \eq{e:alfvenr2} is likely to hold nearly until mass transfer, meaning there is always sufficient time to set up an equilibrium field capable of stably transmitting angular momentum throughout the \wdw{} and enforcing solid-body rotation. \Eq{e:windupr2} is more restrictive, however, and shows that even for a fossil field of initial magnitude of $\Bfoss\sim10^6$ G  in a \cowd{} (\se{s:reslockval}), excessive wind up may begin to occur below an orbital period of $\sim1$ min. Nonetheless, at orbital periods of $\sim100$ min where standing-wave resonance locks occur, which require solid-body rotation, \eq{e:windupr2} requires only modest fields of $\sim200$ G.

Lastly, although angular momentum transport occurs due to a wound-up equilibrium magnetic field, torsional Alfv\'en waves can also be excited, which cause oscillations in the differential rotation profile. Formally, a phase-mixing timescale must elapse before such waves damp \citep{spruit99}. However, due to the slowly varying torque, we expect Alfv\'en wave excitation to be weak. Indeed, in simulations of the solar magnetic field's evolution, \citet{charbonneau93} found that Alfv\'en wave amplitudes (in terms of $\delta\Omegsp/\Omegsp$) were small.

\subsection{Transport during an initial resonance lock}\label{e:transirl} 
When a standing wave resonance lock first begins to takes hold, angular momentum is applied exclusively to a thin layer near the outer wave turning point where wave damping predominantly occurs. Since no equilibrium field state has yet developed in this situation, a concern exists that the layer will rapidly synchronize and destroy the lock before it begins (\se{s:reslockval}).

To this end, we first compare the spin-up timescale of the layer to the Alfv\'en travel time $\tA$ across it. Using the torque for a resonance lock from \eq{e:reslock2} and taking $\tA = H/\vA$, where $H$ is the thickness of the layer, we have
\begin{equation}
 \frac{(2/3)4\pi\rho r^4 H \Omegsp}{\Is\Omegsp/\tgw} \gg \frac{H\sqrt{4\pi\rho}}{B_r}.
\end{equation}
Evaluating $\rho$ at the radiative-convective boundary (RCB), letting $B_r\sim\Bfoss$, and setting $r\approx R$, this becomes
\begin{equation}\label{e:alfvenlocr1}
 \Bfoss \gg \sqrt{\frac{4\pi}{\rho_\mr{rcb}}}\left(\frac{3\Is}{8\pi R^4\tgw}\right).
\end{equation}
We can evaluate this further as
\begin{equation}\label{e:alfvenlocr2}
 \Bfoss\gg10^{-3}\ \mr{G}\;\left( \frac{\Porb}{200\ \mr{min}} \right)^{-8/3} \FC,
\end{equation}
where
\begin{equation*}
 \FC = \left( \frac{\rho_\mr{rcb}}{8.4\times10^{-7} \rho_\mr{c}} \right)^{-1/2}\left( \frac{M'}{M} \right)\left( \frac{1+M'/M}{2} \right)^{-1/3}\left( \frac{\Is}{0.18MR^2} \right)\left(\frac{M}{0.60\Msun} \right)^{13/6}\left( \frac{R}{0.013\Rsun} \right)^{-1/2}
\end{equation*}
and $\rho_\mr{c}$ is the central density.

Secondly, even if no critical layer occurs, it is still necessary for the \emph{global,} wound-up equilibrium field to develop quickly in order for an initial resonance lock to develop. Specifically, in order for a particular resonance to halt the increase in the tidal driving frequency $\omegt=2(\Omegdiff)$, its locally applied tidal torque must be communicated globally  fast enough relative to the timescale over which the torque changes significantly. If this cannot occur, the system sweeps through the resonance without locking (see also \se{s:rtunnel}). The timescale over which a maximally resonant torque decays by roughly a factor of two is given by the time for the detuning $\detun$ to increase from zero to of order the associated mode's damping rate $\gamma_n$ (see e.g.\ \eqp{e:torque2}; also \app{a:lindamp}). Comparing to the global Alfv\'en travel time $\langle t_\mr{A}\rangle$, this condition becomes
\begin{equation}\label{e:rlockr1} 
 \Bfoss \gg \frac{\Omega}{\gamma_n\,\tgw}\int_0^R \sqrt{4\pi\rho}\;dr,
\end{equation}
which is identical to \eq{e:alfvenr1} except with an additional factor of $\Omega/\gamma_n$ on the right-hand side. Evaluating further, we have
\begin{equation}\label{e:rlockr2}
 \Bfoss \gg 100\ \mr{G}\;\left( \frac{\Porb}{200\ \mr{min}} \right)^{-11/3}\left( \frac{\gamma_n^{-1}}{120\ \mr{yr}} \right)\FA,
\end{equation}
where $\FA$ was defined in \app{a:sbody}. The value of $100$ G reduces to $6$ G for a \hewd{} (\heid7 from \ta{t:wdmods}), holding $\gamma_n$ constant. This is a much stricter requirement than \eq{e:alfvenlocr2}, but still seems likely to be satisfied given typical \wdw{} fields (\se{s:reslockval}).

\section{Global normal mode analysis}\label{a:globlin} 
\subsection{Mode dynamics}\label{a:modedyn} 
Here we give an overview of linear normal mode analysis as it applies to tidal interactions.  As such, we assume the fluid motions generated by the tidal potential represent standing waves; we discuss the possibility of traveling waves in \se{s:trav}.

In linear perturbation theory, we expand all fluid variables in spherical harmonics angularly, indexed by $l$ and $m$, and adiabatic normal modes radially, indexed by the number of radial nodes $n$, as
\begin{equation}\label{e:modeexp1} 
\delta X(r,\theta,\phi,t) = \sum_{l=2}^\infty \sum_{m=-l}^l \sum_n q_{nlm}(t) \delta X_{nlm}(r)Y_{lm}(\theta,\phi).
\end{equation}
We computed normal modes using the ADIPLS stellar pulsation package \citep{dalsgaard08}. In this work we concern ourselves only with quadrupolar eigenmodes, and consider only circular orbits, meaning we can let $l=2$ and $m=\pm 2$, since $m=0$ modes have no time dependence with zero eccentricity. Thus \eq{e:modeexp1} becomes
\begin{equation}\label{e:modeexp2} 
\delta X(r,\theta,\phi,t) = 2\sum_n\mr{Re}\left[ q_{n}(t) \delta X_{n}(r) Y_{22}(\theta,\phi)\right],
\end{equation}  
with $m=2$ used throughout.

We can write the momentum equation schematically as \citep{press77}
\begin{equation}\label{e:momeq} 
\rho \ddot{\vecxi}  =-\rho\mathcal{L}[\vecxi] -\rho\nabla U,
\end{equation} 
where the linear internal acceleration operator $\mathcal{L}$ satisfies $\mathcal{L}[\vecxi_n]=\omega_n^2\vecxi_n$, $\omega_n$ is an eigenfrequency, and $U$ is the tidal potential. The corotating frame mode amplitude equations determining the behavior of each $q_n$ can be obtained from \eq{e:momeq} by taking the scalar product with $\vecxi_n^*$ on both sides and integrating over the star, yielding (e.g.\ \citealt{weinberg12})
\begin{equation}\label{e:modeamp} 
\ddot{q}_n + 2\gamma_n \dot{q}_n + \omega_n^2 q_n = 2 \omega_n^2 \epsilon W Q_n e^{-i\omegt t},
\end{equation} 
where $W=\sqrt{3\pi/10}$, $\epsilon=(M'/M)(R/a)^3$ is the tidal factor, $\omegt=m(\Omega-\Omegsp)$ is the tidal driving frequency, and the linear overlap integral $Q_n$ is discussed in \app{a:linover} \citep{press77}. We have also inserted into \eq{e:modeamp} a damping rate $\gamma_n$, the calculation of which we describe in \app{a:lindamp}. We normalize our normal modes by setting
\begin{equation}\label{e:norm} 
\Es = E_n = \int_0^R 2\omega_n^2\left(\xi_{r,n}^2+l(l+1) \xi_{\h,n}^2\right)\rho r^2dr.
\end{equation}

Given slowly varying orbital and stellar properties, the steady-state solution to \eq{e:modeamp} is
\begin{equation}\label{e:qsolst} 
q_n(t) = 2\,\epsilon\, Q_n W \left(\frac{\omega_n^2}{(\omega_n^2-\omegt^2)-2i\gamma_n\omegt}\right) e^{-i\omegt t}.
\end{equation} 
The above amplitude applies in the corotating stellar frame; in the inertial frame the time dependence $e^{-i\omegt t}$ instead becomes $e^{-im\Omega t}$.

\subsection{Angular momentum and energy transfer}\label{a:torque} 
Assuming a circular orbit and alignment of spin and orbital angular momenta, the secular quadrupolar tidal torque on a star is given by an expansion in quadrupolar ($l=2$) normal modes as (\citealt{burkart12} Appendix C1 and references therein)
\begin{equation}\label{e:gentorque} 
\torq = 8m \Es\epsilon^2 W^2\sum_n Q_n^2\frac{\omega_n^2\omegt\gamma_n}{(\omega_n^2-\omegt^2)^2+4\omegt^2\gamma_n^2},
\end{equation} 
where most variables are defined in the previous section.

The tidal energy deposition rate into the star $\Edtide$ can be determined from $\torq$ by the relation
\begin{equation}\label{e:etau} 
-\dot{E}_\mr{orb}=\Edtide = \Omega\,\torq,
\end{equation} 
valid only for a circular orbit. This can be derived by differentiating standard equations for the energy and angular momentum of a binary with respect to time, setting $\dot{e}=e=0$, and noting that the tidal torques and energy deposition rates in each star of a binary are independent.

The total energy contained in the linear tide can be expressed in the corotating frame as \citep{schenk02}
\begin{equation}
\Etide = \frac{1}{2}\left\langle\dot{\vecxi},\dot{\vecxi}\right\rangle + \frac{1}{2}\left\langle\vphantom{\dot{\vecxi}}\vecxi,\mathcal{L}[\vecxi]\right\rangle,
\end{equation} 
where the operator $\mathcal{L}$ was introduced in \app{a:modedyn}. Using results from \app{a:modedyn}, we can evaluate this expression as
\begin{equation}
\Etide =  2\Es\epsilon^2W^2\sum_n Q_n^2\frac{\omega_n^2(\omegt^2 + \omega_n^2)}{(\omega_n^2-\omegt^2)^2+4\omegt^2\gamma_n^2}.
\end{equation} 
Lastly, since the great majority of the tidal energy is in the lowest-order modes, i.e.\ the equilibrium tide, which satisfy $\omega_n\gg\omegt$, we can further set $\omegt\approx0$ in the previous equation to derive the simple expression
\begin{equation}\label{e:Etidesimp} 
\Etide \approx \lambda\Es\epsilon^2,
\end{equation}
where
\begin{equation}
 \lambda = 2W^2\sum_n Q_n^2.
\end{equation} 

Several limits can be taken of our general torque expression in \eq{e:gentorque}. First, if we assume $\omega_n \gg (\omegt,\gamma_n$), we arrive at the equilibrium tide limit:
\begin{equation}
\torq_\mr{eq} = 8m\Es\epsilon^2 W^2\sum_n Q_n^2\left(\frac{\omegt\gamma_n}{\omega_n^2}\right).
\end{equation}
\citet{willems10} showed that in linear theory  the equilibrium tide provides a negligible torque due to the very weak damping present in \wdws{}, resulting from their high densities and long thermal times. We confirm this result: e.g.,
the tidal quality factor associated with damping of the quadrupolar equilibrium tide in \wdws{} is (\eqp{e:Qtorq})
\begin{equation}
 \Qt^\mr{eq} = \frac{1}{8}\frac{\sum_n Q_n^2}{\sum_n Q_n^2 \omegt\gamma_n/\omega_n^2}\gtrsim 10^{13},
\end{equation}
which is much larger than the effective value of $\Qt$ for the dynamical tide determined in \se{s:Q}. The resonant dynamical tide ($\omega_n\sim\omegt$ in \eqp{e:gentorque}) is discussed in \S\S\ \ref{s:reslock} \& \ref{s:trav}.

\subsection{Damping}\label{a:lindamp}

\begin{figure*}\centering
\begin{tabular}{lr}
  \begin{overpic}{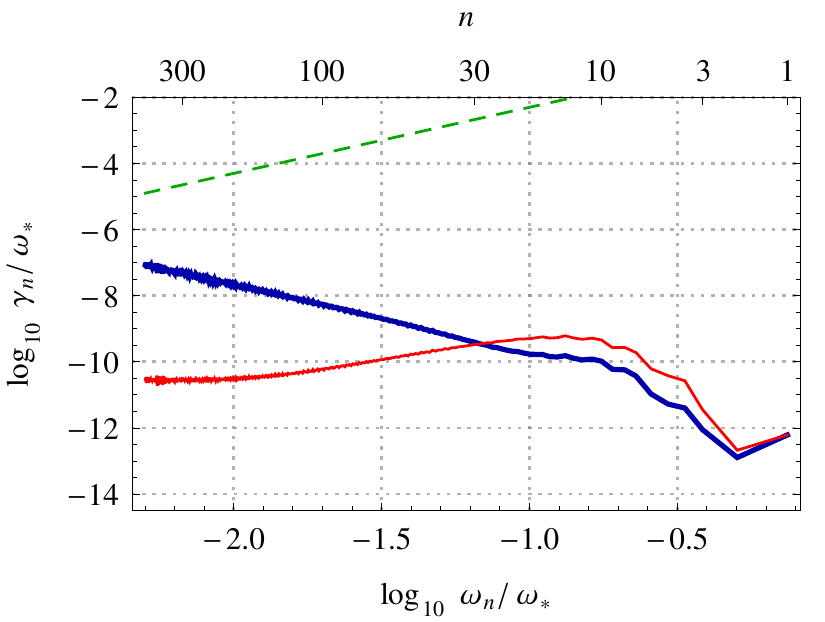}
  \put(18,17){\transparent{.8}\colorbox{white}{\normalsize\transparent{1}\heid7: $0.2\Msun,\ \Teff = 7,000$ K}}
  \put(25,45){\transparent{.8}\colorbox{white}{\transparent{1}$\gamma^\mr{diff}_n$}}
  \put(46,27.5){\transparent{.8}\colorbox{white}{\transparent{1}$\gamma^\mr{turb}_n$}}
  \put(50,56){\transparent{.8}\colorbox{white}{\normalsize\transparent{1}$\alpha_n$}}
  \put(88,55){\colorbox{white}{\Large 1}}
  \end{overpic}
&  
  \begin{overpic}{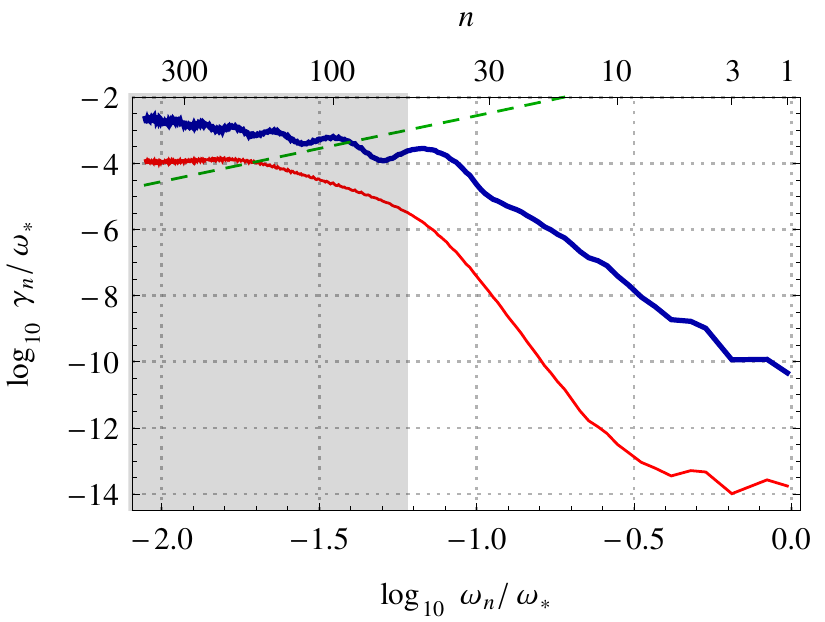}
  \put(18,17){\transparent{.8}\colorbox{white}{\normalsize\transparent{1}\heid{10}: $0.2\Msun,\ \Teff =9,900$ K}}
  \put(88,55){\colorbox{white}{\Large 2}}
  \end{overpic}\\
  \begin{overpic}{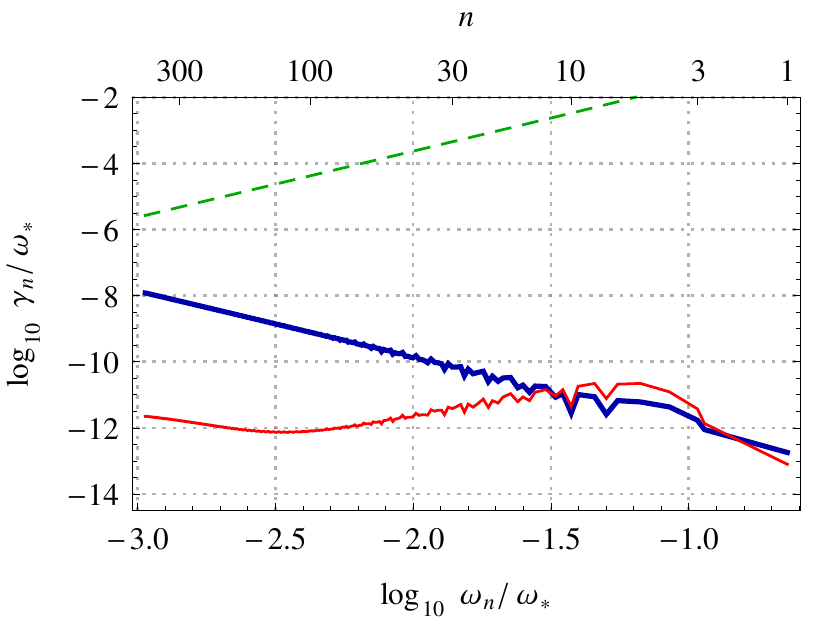}
  \put(18,17){\transparent{.8}\colorbox{white}{\normalsize \transparent{1}\coid{6}: $0.6\Msun,\ \Teff =5,500$ K}}
  \put(88,55){\colorbox{white}{\Large 3}}
  \end{overpic}
  &
  \begin{overpic}{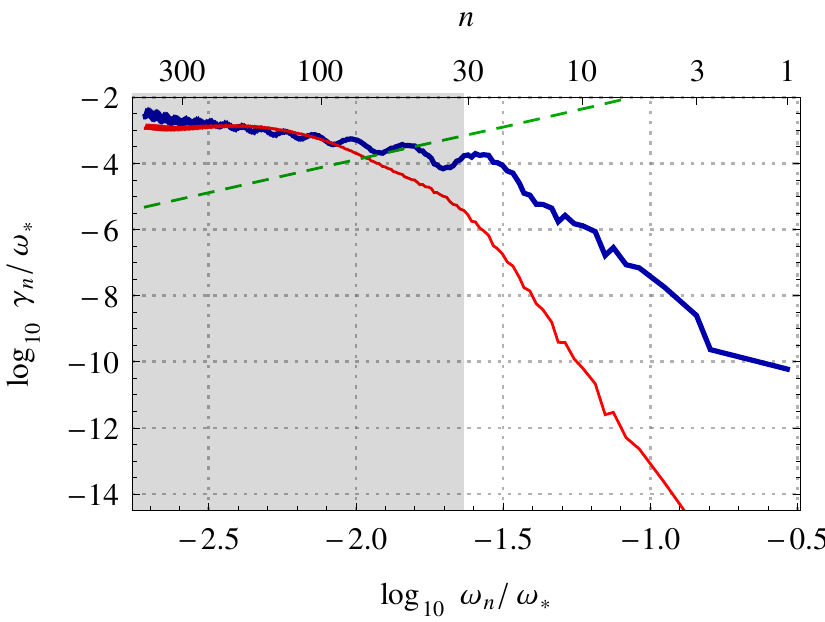}
  \put(18,17){\transparent{.8}\colorbox{white}{\normalsize\transparent{1}\coid{12}: $0.6\Msun,\ \Teff =12,000$ K}}
  \put(88,55){\colorbox{white}{\Large 4}}
  \end{overpic}
\end{tabular}
  \caption{Plots of the contributions to eigenmode damping rates due to thermal diffusion $\gamma_n^\mr{diff}$ (thick blue lines) and turbulent convection $\gamma_n^\mr{turb}$ (red lines), as well as the effective damping rate for traveling waves $\alpha_n=2\pi/\tgroup{}_{,\,n}$ (dashed green lines), as functions of the eigenmode frequency $\omega_n$ (in units of the dynamical frequency $\omegdyn$) and radial order $n$, for the first 400 g-modes in four of our fiducial \wdw{} models (\ta{t:wdmods}). We describe the computation of these quantities in \app{a:lindamp}. Inside the shaded regions in panels 2 \& 4, g-modes become traveling waves due to rapid thermal diffusion near their outer turning points. In this case, approximating wave damping rates using $\gamma_n^\mr{diff} + \gamma_n^\mr{turb}$ is invalid, and a traveling wave formalism such as that discussed in \se{s:trav} must be used. Note that the \wdw{} models shown here are different from those in \fig{f:Qplots}.}
  \label{f:gammaplots}
\end{figure*}

We consider two different damping processes in this appendix: thermal diffusion and turbulent convection. We also calculate g-modes' group travel times, which sets the effective damping time for traveling waves. \fig{f:gammaplots} shows plots of all three of these quantities for several of our fiducial \wdw{} models, and demonstrates that thermal diffusion is the dominant source of damping for high-order g-modes in \wdw{}s.

Damping due to electron conduction and radiative diffusion can be estimated simultaneously as \citep{goodman98}
\begin{equation}\label{e:damp} 
\gamma_n^\mr{diff} = \frac{1}{2E_n}\int \chi k_r^2\frac{dE_n}{dr}dr,
\end{equation} 
where $dE_n/dr$ is the integrand of \eq{e:norm}, $\chi$ is the thermal diffusivity (including both radiative diffusion and electron conduction), and $k_r$ is the radial wavenumber. In terms of an effective opacity $\kappa$, $\chi$ can be expressed as
\begin{equation}
 \chi = \frac{16 \sigma T^3}{3\kappa \rho^2 c_p}.
\end{equation}
For high-order g-modes, $k_r$ is given by \citep{dalsgaard08}
\begin{equation}\label{e:kr} 
k_r^2 = k_\h^2\left( \frac{N^2}{\omega^2}-1 \right),
\end{equation}
where $ k_\h^2 = l(l+1)/r^2$ is the angular wavenumber.
The integration in \eq{e:damp} is performed up to the adiabatic cutoff radius defined by $\omega\, \ttherm=1$ \citep{unno89}, where $\ttherm=pc_pT/gF$ is the local thermal time. A wave's group travel time across a scale height must remain smaller than its local damping time in order for the wave to reflect. This criterion will always be broken for sufficiently long-period waves since the radial wavenumber $k_r$ grows as $\omega^{-1}$; nonetheless, we find that thermal diffusion is never strong enough to invalidate the standing wave assumption for all of the modes used in this work that are capable of effecting standing wave resonance locks.

To estimate the turbulent convective damping rate, we rely on the calibration of convective viscosity performed by  \citet{penev09}. The formula we employ is
\begin{equation}
 \gamma_n^\mr{turb} \sim \frac{\omega_n^2}{E_n}\int \rho r^2 \nu_\mr{turb} \left[ s_{0'}\left(\frac{d\xir}{dr}\right)^2 + s_1l(l+1)\left( \frac{d\xih}{dr} \right)^2 \right]dr,
\end{equation}
where $s_{0'}=0.23$ and $s_1=0.084$ \citep{penev11}. For the effective turbulent viscosity $\nu_\mr{turb}$, we use equation~(11) of \citet{shiode12}:
\begin{equation}
 \nu_\mr{turb} = L v_\mr{conv} \, \mr{min}\left[  \frac{1}{\Pi_\mr{min}}\left( \frac{2\pi}{\omega\, \teddy} \right)^2,\left( \frac{2\pi}{\omega \,\teddy} \right),\Pi_\mr{max}\right],
\end{equation}
where $v_\mr{conv}$ is the convective velocity, $L$ is the mixing length,  $\teddy=v_\mr{conv}/L$, $\Pi_\mr{min} = 0.1$, and $\Pi_\mr{max} = 2.4$.

Lastly, the effective damping rate applicable in the traveling wave regime is the inverse group travel time $\alpha = 2\pi/\tgroup$; we can calculate $\tgroup$ as
\begin{equation}
 \tgroup = 2\int \frac{dr}{|v_\mr{group}|},
\end{equation}
where $v_\mr{group}=d\omega/dk_r$ and the integration is over the propagation cavity where the wave frequency $\omega<N$. Using the dispersion relation from \eq{e:kr}, this becomes
\begin{equation}\label{e:alpha} 
 \alpha = \pi\left( \int \frac{k_\h\,N}{\omega^2}dr \right)^{-1}.
\end{equation} 
For $\omega\ll N$, $\alpha\propto \omega^{2}$; this proportionality is verified in \ta{t:eigprops} and \fig{f:gammaplots}.

\subsection{Linear overlap integral}\label{a:linover}
The linear overlap integral for quadrupolar eigenmodes, introduced in \eq{e:modeamp}, can be expressed as
\begin{equation}\begin{split}\label{e:overlap}
Q_n&=\frac{1}{M R^l}\int_0^{R} l\left(\xi_{r,n}+(l+1)\xi_{\h,n}\right)\rho r^{l+1} dr \\
&= \frac{1}{M R^l}\int_0^{R} \delta\rho_{n}\, r^{l+2} dr\\
& = -\frac{R}{GM}\cdot\frac{2l+1}{4\pi}\cdot\delta\phi_n(R),
\end{split}\end{equation}
with $l=2$, where $\xih$ is the horizontal fluid displacement. The second equality can be derived by substituting the continuity equation, and the third equality by substituting Poisson's equation. However, all three of these methods of calculating $Q_n$ suffer from numerical difficulties, presumably arising due to a failure of orthogonality or completeness of the numerically computed eigenmodes \citep{fuller11}, or to small inconsistencies in the stellar model \citep{fuller12}.

Thus we now consider a more stable way of numerically evaluating $Q_n$, which is what we actually employed in our calculations and the fits in \ta{t:eigprops}. We again focus on $l=2$ modes, but the technique can easily be extended to arbitrary $l$. First, the tidally generated displacement field is given as a sum of normal modes in \eq{e:modeexp1}. If we set the tidal driving frequency $\omegt$ and the damping rate $\gamma_n$ to zero in \eq{e:qsolst} and substitute into \eq{e:modeexp2} (while keeping the tidal factor $\epsilon$ nonzero), we recover the equilibrium tide limit. However, the equilibrium tide can alternatively be obtained by directly solving the inhomogeneous linear stellar oscillation equations in the zero-frequency limit; see e.g.\ \citet{weinberg12} Appendix A.1. Equating these two alternate expressions, taking the scalar product with $\rho \vecxi_{n}^*$ on both sides, integrating over the star, and solving for $Q_{n}$, one obtains
\begin{equation}
 Q_{n} = \frac{\omega_{n}^2 }{W\Es}\int_0^R\left( X_\mr{eq}^{r} \xi_{n}^r + l(l+1)X^\h_\mr{eq} \xi_{n}^\h  \right)\rho r^2 dr,
\end{equation}
where
\begin{equation}
 \vec{X}_\mr{eq}(r,\theta,\phi) = \sum_{m=\pm2} \left(X_\mr{eq}^{r}(r) \hat{r} + rX_\mr{eq}^\h(r)\svec{\nabla}\right)Y_{2m}(\theta,\phi)
\end{equation} 
is the numerically computed $l=2$ equilibrium tide scaled to $\epsilon=1$.

Finally, there is one further method of computing $Q_n$ that we have employed, which also uses a solution to the inhomogeneous equations. In this method, however, instead of comparing alternate computations of the equilibrium tide, we instead evaluate the inhomogeneous tidal response very near an eigenfrequency. The overlap can then be extracted by fitting the resulting Lorentzian profile of e.g.\ the tidal energy. This method numerically agrees very well with the equilibrium tide method described above.

\section{Verification of traveling wave torque approximation}\label{a:travver}
Here we will justify our traveling wave tidal torque approximation described in \se{s:trav_intro} using several distinct lines of reasoning. Our goal is to explain why the traveling wave torque can be expressed in terms of the properties of global eigenmodes as in \eq{e:travtorque}. This is a different approach than is typical in the literature.

First, in the limit that the wave damping time is much longer than the group travel time, the tidal response is well approximated as a standing wave. Thus taking the standing wave torque and setting the damping rate equal to the inverse group travel time (\app{a:lindamp}) represents a natural method of smoothly transitioning to the traveling wave limit, since it corresponds to the situation where a wave is nearly completely absorbed over one travel time. One apparent difficulty with the resulting expression in our \eq{e:travtorque} is that it appears to contain explicit dependence on the wave travel time, which seems paradoxical, since a traveling wave has no information about the extent of the propagation cavity. This is, however, simply an artifact of our normalization convention. To show this, we first note that our approximation for $\torqtrav$ depends on $\alpha=2\pi/\tgroup$ (\app{a:lindamp}) only through
\begin{equation}\label{e:travtorqdep} 
\torqtrav\propto Q^2/\alpha.
\end{equation} 
Next, given the appropriate WKB expression for $\xih$ \citep{dalsgaard08},
\begin{equation}
 \xih \approx A\sqrt{\frac{N}{\rho r^{3} \Lambda\omegt^{3}}} \sin\left(\int k_r dr + \delta\right),
\end{equation}
where $A$ is a constant, we impose our normalization convention from \eq{e:norm} and use the fact that $\xih\gg\xir$ for g-modes to obtain
\begin{equation}\begin{split}\label{e:normapprox}
 \Es &\approx 2\omegt^2\int\Lambda^2\xih^2 \rho r^2dr\\
 &\approx A^2\left( \frac{\pi\omegt \Lambda}{\alpha} \right),
\end{split}\end{equation}
having set $\sin^2\rightarrow 1/2$. This implies that unnormalized eigenfunctions (which are independent of global integrals) must be multiplied by $A\propto \alpha^{1/2}$ in order to be normalized properly. Since the overlap integral $Q$ is linear in the eigenfunctions (\app{a:linover}), re-examining \eq{e:travtorqdep} shows that $\torqtrav$ is indeed independent of $\alpha$ and hence $\tgroup$.

An alternate justification of our traveling wave torque expression can be obtained by considering how a traveling wave of a given frequency $\omegt$ can be expressed in terms of global standing wave eigenmodes, which form a complete basis  \citep{dyson79}. Obtaining a traveling wave functional form of $e^{i(kr\pm\omegt t)}$ requires summing at least two real-valued eigenmodes with a relative global phase difference between their complex amplitudes of $\pm\pi/2$, where both modes possess frequencies close to $\omegt$. Examining \eq{e:qsolst}, we see that the phase of a standing mode's amplitude is given by $\arctan(\gamma/\detun)$. Thus, in order to approximate a traveling wave of commensurate frequency, two adjacent eigenmodes 1 and 2 must satisfy
\begin{equation}
 \arctan\left(\frac{\gamma_1}{\detun_1}\right) - \arctan\left(\frac{\gamma_2}{\detun_2}\right) = \pm \frac{\pi}{2},
\end{equation}
which simplifies to $\gamma_1\gamma_2 = \detun_1\detun_2$. Setting $|\detun_{1,2}| \approx (\Delta P_0/2\pi)\omegt^2 \approx \alpha$, where $\Delta P_0$ is the asymptotic g-mode period spacing, we find that the damping rate required to produce a traveling wave is $\gamma\sim 2\pi/\tgroup$, consistent with our \eq{e:travtorque}.

\begin{figure}\centering
\includegraphics{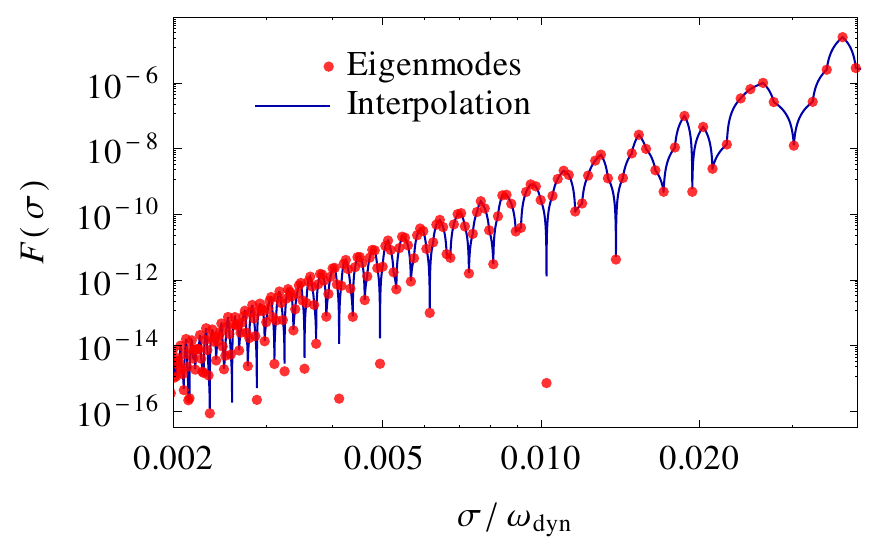}\includegraphics{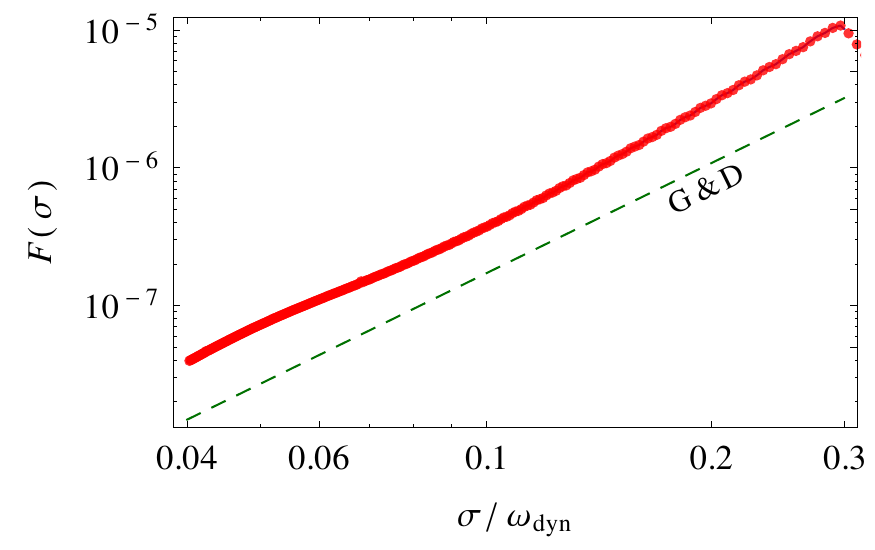}
  \caption{Dimensionless traveling wave tidal torque $F(\omegt)=\torqtrav/\epsilon^2\Es$, computed as described in \se{s:trav_intro}. Red points show direct evaluations of \eq{e:travtorque} at eigenmode frequencies, while blue lines are linear interpolations of these values. \emph{Left panel:} Results for our $0.6\Msun$, $\Teff=5,500$ K \coid6 \wdw{} model (\ta{t:wdmods}). To facilitate straightforward comparison, this plot employs the same axes and conventions as in Figure 8 of \citet{fuller12}. \emph{Right panel:} Results for a solar model, which agrees reasonably well with the semi-analytic traveling wave result in equation (13) of \citet{goodman98} (dashed green line).}
  \label{f:litcomp}
\end{figure}

Lastly, we will quantitatively demonstrate that our approximation for the traveling wave tidal torque  reproduces results available in the literature. First, the left panel of \fig{f:litcomp} shows our traveling wave torque evaluated for our \coid6 WD model (\ta{t:wdmods}), expressed in the dimensionless form $F(\omegt) = \torqtrav/\epsilon^2\Es$, as a function of the $l=m=2$ tidal forcing frequency $\omegt=2(\Omegdiff)$. Comparing this to Figure 8 of \citet{fuller12}, which employs the same conventions and axes, shows reasonable agreement. In particular, both exhibit the jagged variation with frequency discussed in \se{s:excite}. Our result has a slightly steeper overall trend, leading to a smaller torque at low frequencies; however, because excitation is sensitive to the details of the composition boundaries in the model, there is no reason to expect detailed agreement. In addition, the right panel of \fig{f:litcomp} compares traveling wave results for solar-type stars from \citet{goodman98} (using their equation 13) with our \eq{e:travtorque} applied to a solar model. Both possess the same power-law scaling with frequency, and agree in normalization within a factor of $\sim2$.

\section{White dwarf models}\label{a:models}
We used MESA version 4298 \citep{paxton11} to produce our \hewd{}  models (\ta{t:wdmods}: \heid5, \heid7, and \heid{10}). We used three inlists. The first evolves a $1.6\Msun$ star with $Z=0.02$ from ZAMS to where $0.198\Msun$ of its core is  locally at least 90\% helium. Salient non-default parameter values are:
\begin{align*}
\text{\texttt{mesh\_delta\_coeff}} &\ \ =\ \  \text{\texttt{0.5}}\phantom{\text{\texttt{mesh\_delta\_coeff}}}\\
\text{\texttt{h1\_boundary\_limit}} &\ \ =\ \   \text{\texttt{0.1}}\\
\text{\texttt{h1\_boundary\_mass\_limit}}  &\ \ =\ \   \text{\texttt{0.198}}
\end{align*}
The second smoothly removes the outer $1.4\Msun$, leaving a hydrogen layer with $\sim1$\% of the remaining mass. This is achieved with:
\begin{align*}
\text{\texttt{relax\_mass}} &\ \ =\ \  \text{\texttt{.true.}} \\
\text{\texttt{new\_mass}} &\ \ =\ \  \text{\texttt{0.2}}
\end{align*}
The third evolves the resulting $0.2\Msun$ \wdw{} until $\Teff=5,000$~K. We invoke MESA's element diffusion routine even where the plasma interaction parameter $\Gamma>1$. Salient non-default parameter choices are:
\begin{align*}
\text{\texttt{mesh\_delta\_coeff}} &\ \  =\ \  \text{\texttt{0.05}} \\
\text{\texttt{which\_atm\_option}} &\ \  = \ \   \text{\texttt{`simple\_photosphere'}} \\
\text{\texttt{surf\_bc\_offset\_factor}} &\ \  = \ \   \text{\texttt{0}} \\
\text{\texttt{do\_element\_diffusion}} &\ \  = \ \   \text{\texttt{.true.}} \\
\text{\texttt{use\_Ledoux\_criterion}} &\ \  = \ \   \text{\texttt{.true.}} \\
\text{\texttt{diffusion\_gamma\_full\_off}} &\ \  =\ \   \text{\texttt{1d10}} \\
\text{\texttt{diffusion\_gamma\_full\_on}} &\ \  = \ \   \text{\texttt{1d10}} \\
\text{\texttt{diffusion\_T\_full\_off}} &\ \  =  \ \  \text{\texttt{-1}} \\
\text{\texttt{diffusion\_T\_full\_on}} &\ \  = \ \   \text{\texttt{-1}} \\
\text{\texttt{diffusion\_Y\_full\_off}} &\ \  = \ \   \text{\texttt{-1}} \\
\text{\texttt{diffusion\_Y\_full\_on}} &\ \  =  \ \  \text{\texttt{-1}}
\end{align*}

Our \cowd{} models (\ta{t:wdmods}: \coid6 and \coid{12}) were produced by solving for hydrostatic equilibrium subject to heat transport by radiative diffusion and electron conduction \citep{hansen04}. We use the OPAL EOS and effective opacities \citep{rogers96} in the \wdw{} outer layers, and transition to the Potekhin-Chabrier EOS and electron conduction opacities \citep{potekhin10} where the electrons begin to become degenerate. We use a mixture of 25\% carbon, 75\% oxygen for the inner 98\% of the model's mass, then add a helium layer with 1.7\% of the mass, and finally a hydrogen layer with the remaining 0.17\%. We smooth the composition transition regions over $\sim0.1H_p$ with a Gaussian profile, where $H_p$ is a pressure scale height. We treat convection with mixing length theory, using $L=H_p$ for the mixing length.

\end{appendix}

\bibliography{wdinsp}

\end{document}